\documentclass[12pt]{article}
\pdfoutput=1
\usepackage{subfigure}
\usepackage{amssymb,amsmath}
\usepackage{graphicx}
\usepackage{color}
\usepackage[colorlinks=true
,urlcolor=blue
,citecolor=blue
,linkcolor=blue
,pagecolor=blue
,linktocpage=true
,pdfproducer=medialab
]{hyperref}
\usepackage[a4paper,width=15.2cm]{geometry}
\makeatletter \renewcommand{\@dotsep}{10000} \makeatother

\begin{document}

\title{{\bf Higgs and Sparticle Spectroscopy with Gauge-Yukawa Unification}
}

\date{}
\maketitle
\vspace{-1.75cm}
\begin{center}
{\large
Ilia Gogoladze
\footnote{
Email: ilia@bartol.udel.edu. On leave of absence from:
Andronikashvili Institute of Physics, GAS, Tbilisi, Georgia.},
Rizwan Khalid
\footnote{
Email: rizwan.hep@gmail.com. On study leave from:
Centre for Advanced Mathematics \& Physics of the National
University of Sciences \& Technology, H-12, Islamabad, Pakistan. },
Shabbar Raza
\footnote{
Email: shabbar@udel.edu. On study leave from:
Department of Physics, FUUAST, Islamabad, Pakistan.}
and
Qaisar Shafi
}

\vspace{0.75cm}

{\it
Bartol Research Institute, Department of Physics and Astronomy, \\
University of Delaware, Newark, DE 19716, USA 
}

\section*{Abstract}

\end{center}

We explore the Higgs and sparticle spectroscopy of 
supersymmetric $SU(4)_c \times SU(2)_L \times SU(2)_R$ models in which the 
three MSSM gauge couplings and third family ($t$-$b$-$\tau$) Yukawa 
couplings are all unified at $M_{\rm GUT}$. This class of models can be
obtained via compactification of a higher dimensional theory. 
Allowing for opposite sign gaugino masses 
and varying $m_t$ within $1\sigma$ of its current central value yields a 
variety of gauge-Yukawa unification as well as WMAP compatible neutralino dark matter solutions. 
They include mixed bino-Higgsino dark matter, stau and gluino 
coannihilation scenarios, and the A-resonance solution. 

\thispagestyle{empty}
\setcounter{page}{0}
\newpage
\hrule
\bigskip
\tableofcontents
\bigskip\bigskip
\hrule

\section{Introduction}

Supersymmetric (SUSY) $SO(10)$ GUT (grand unified theory), in contrast to its 
non-SUSY version, yields third family ($t$-$b$-$\tau$) Yukawa 
unification via the unique
renormalizable Yukawa coupling $16 \cdot 16 \cdot 10$, where the
10-plet is assumed to contain the two minimal supersymmetric standard
model (MSSM) Higgs doublets $H_u$ and $H_d$, and the 16-plet contains
the 15 chiral fermions per family of the standard model (SM) as
well as the right handed neutrino. The implications of this unification
have been extensively explored over the years
\cite{big-422,bigger-422}. More recently, it has been argued in
\cite{Baer:2008jn, Gogoladze:2009ug} that $SO(10)$ Yukawa unification
predicts relatively light ($\lesssim$ TeV) gluinos, which can be readily
tested \cite{Baer:2009ff} at the Large Hadron Collider (LHC). The
squarks and sleptons  turn out to have masses in the multi-TeV range.
Moreover, it is argued in \cite{Baer:2008jn, Gogoladze:2009ug} that
the lightest neutralino is not a viable cold dark matter candidate,
at least in the simplest models of $SO(10)$ Yukawa unification.

Spurred by these developments we have investigated $t$-$b$-$\tau$ Yukawa
unification \cite{Gogoladze:2009ug, Gogoladze:2009bn,Gogoladze:2010fu} in the
framework of supersymmetric $SU(4)_c \times SU(2)_L \times SU(2)_R$ \cite{pati}
(4-2-2, for short). The 4-2-2 structure allows us to
consider non-universal gaugino masses while retaining Yukawa
unification. An important conclusion
reached in \cite{Gogoladze:2009ug, Gogoladze:2009bn} is that with
same sign non-universal gaugino soft terms, 
Yukawa unification in 4-2-2 is compatible
with neutralino dark matter, with gluino co-annihilation
\cite{Gogoladze:2009ug, Gogoladze:2009bn, Profumo:2004wk, Ajaib:2010ne}
playing an important role.
By considering opposite sign gauginos with {$\mu<0,M_2<0,M_3>0$} 
(where $\mu$ is the bilinear Higgs mixing term, and $M_2$ and $M_3$ are 
the soft supersymmetry breaking gaugino mass terms corresponding respectively to 
$SU(2)_L$ and $SU(3)_c$)  in
\cite{Gogoladze:2010fu} we have shown that Yukawa coupling
unification consistent with known experimental constraints
is realized in 4-2-2. With $\mu<0$ and opposite sign gauginos, Yukawa coupling
unification is achieved for $m_0 \gtrsim 300\, {\rm GeV}$, as opposed
to $m_0 \gtrsim 8\, {\rm TeV}$ for the case of same sign gauginos,
by taming the finite corrections to
the b-quark mass. By considering gauginos with $M_2 <0$ and $M_3>0$ 
and $\mu<0$, we can obtain the correct sign for the 
desired contribution to $(g-2)_\mu$ \cite{Bennett:2006fi}.
This enables us to simultaneously satisfy the requirements of  $t$-$b$-$\tau$ Yukawa
unification, neutralino dark matter and $(g-2)_\mu$, as well as a
variety of other known bounds.

Encouraged by the abundance of solutions and coannihilation channels available in
the case of Yukawa unified 4-2-2, it is natural to try to further constrain this model.
One possible way is to impose unification of $t$-$b$-$\tau$ Yukawa couplings 
with the MSSM gauge couplings at $M_{\rm GUT}$. This is partially inspired  
from the observation that at $M_{\rm GUT}$, the unified gauge 
coupling for the MSSM with TeV scale supersymmetry 
is $\sim 0.7$, while the corresponding third generation Yukawa couplings 
are of order $0.6$. This suggests that the origin of Yukawa 
couplings and gauge interaction may be closely related, and indeed 
higher dimensional supersymmetric models have been constructed   
that predict gauge-Yukawa unification (GYU) 
\cite{su8,Burdman:2002se,GHM3}. We will briefly summarize one such model later 
in the paper. The phenomenology of this idea was 
studied in \cite{GHM3},  where it was shown how a   
suitable choice of low scale SUSY threshold corrections can 
yield GYU condition in principle, 
without precisely specifying the origin and values for the soft 
SUSY breaking parameters. 

The main purpose of this paper is to extend the 4-2-2 discussion to
the case of GYU in the framework of gravity 
mediated SUSY breaking scenario. 
In Section \ref{model} we briefly describe the Yukawa unified
4-2-2 model and the boundary conditions
for the soft supersymmetry breaking (SSB) parameters employed in our scan. In Section
\ref{constraintsSection} we summarize the scanning procedure and the
various experimental constraints that we impose. In Section
\ref{muneg} we discuss threshold corrections to the Yukawa
couplings and summarize from previous studies the findings pertaining 
to Yukawa unification. We also present new results in this section for Yukawa unification
for the case of {$\mu>0,M_2>0,M_3<0$} in this section.
In Section \ref{gaugeYukawaSec} we discuss GYU 
with MSSM as the low energy theory.
We first describe a concrete model that breaks to SUSY 4-2-2 at
$M_{\rm GUT}$ and yields gauge-$t$-$b$-$\tau$ Yukawa unification condition.
We then proceed to discuss the role played by threshold corrections to
$\delta y_t$ in order to obtain GYU.
The important role of the top quark mass in implementing GYU
is also emphasized. In Section \ref{results} we present our 
results and highlight
some of the predictions of the GYU 4-2-2 model.
The correlation between direct and
indirect detection of dark matter and the gauge-Yukawa
unification condition is presented in Section \ref{dark} where we also
display some benchmark points. Our conclusions are summarized in
Section \ref{conclusions}.

\section{The 4-2-2 model \label{model}}

In 4-2-2 the 16-plet of $SO(10)$ matter fields consists of
$\psi$ (4, 2, 1) and $\psi_c$ $(\bar{4}, 1, 2)$. The third family Yukawa coupling
$\psi_c \psi H$, where $H(1,2,2)$ denotes the bi-doublet (1,2,2),  yields
the following relation valid at $M_{\rm GUT}$,
\begin{align}
Y_t = Y_b = Y_{\tau} = Y_{\nu_{\tau}}. \label{f1}
\end{align}

In a realistic scenario we can expect corrections to Eq.(\ref{f1}) 
arising, say, from higher dimensional operators. We will assume that 
these are sufficiently small so that Eq.(\ref{f1}) is valid within 
a few percent or so. 

Supplementing 4-2-2 with a discrete left-right (LR) symmetry
\cite{pati,lr} (more precisely C-parity) \cite{c-parity} reduces the
number of independent gauge couplings in 4-2-2 from three to two.
This is because C-parity imposes the gauge
coupling unification condition ($g_L=g_R$) at $M_{\rm GUT}$. We will
assume that due to C-parity the SSB
mass terms, induced at $M_{\rm GUT}$ through gravity mediated
supersymmetry breaking \cite{Chamseddine:1982jx} are equal in magnitude for the  squarks
and sleptons of the three families. The tree level asymptotic MSSM
gaugino SSB masses, on the other hand, can be non-universal from the
following consideration. From C-parity, we can expect that the
gaugino masses at $M_{\rm GUT}$ associated with $SU(2)_L$ and
$SU(2)_R$ are the same ($M_2 \equiv M_2^R= M_2^L$). However, the
asymptotic $SU(4)_c$ and consequently $SU(3)_c$ gaugino SSB masses
can be different. With the hypercharge generator in 4-2-2 given by
$Y=\sqrt{2/5}~(B-L)+\sqrt{3/5}~I_{3R}$, where $B-L$ and $I_{3R}$ are
the diagonal generators of $SU(4)_c$ and $SU(2)_R$, we have the
following asymptotic relation between the three MSSM gaugino SSB
masses:
\begin{align}
M_1=\frac{3}{5} M_2 + \frac{2}{5} M_3. \label{gauginoCondition}
\end{align}

The supersymmetric 4-2-2 model with C-parity thus has two
independent parameters ($M_2$ and $M_3$) in the gaugino sector. In
order to implement Yukawa unification it turns out that the SSB Higgs
mass terms must be non-universal at $M_{\rm GUT}$. Namely,
$m_{H_u}^2<m_{H_d}^2$ at $M_{GUT}$, where $m_{H_u} (m_{H_d})$ is the
up (down) type SSB Higgs mass term. 
Phenomenological studies of the MSSM typically resort to 
`just-so' splitting (see Blazek, Dermisek and Raby in ~\cite{bigger-422}) 
for the MSSM Higgs doublets, while 
remarking in passing that such a splitting may arise, for example, 
from $D$ terms. $D$ terms, however, induce splitting in the 
squarks and sleptons as well. 
It is possible to imagine a simple mechanism to implement 
just-so Higgs SSB mass splitting in either the 4-2-2 or $SO(10)$ models.  

We need Higgs fields other than $H(1,2,2)$ in order to
complete the model. For instance, $\Phi(4,1,2)+\bar\Phi(\bar 4,2,1)$ field
may be used to
break 4-2-2 to the SM gauge group. In the 4-2-2 model, just-so Higgs splitting may be
understood by writing an $SU(2)_R$ violating bilinear term between the
up and down type Higgs doublets.
This may be done, for instance, by considering a super-heavy
Higgs $\Delta$ that transforms as $(1,1,3)$ under 4-2-2. If the MSSM
doublet comes from $H(1,2,2)$ under 4-2-2, the superpotential
will have the interaction $HH\Delta$. We may consider the
SSB trilinear term $A_{\Delta} H H \Delta$ and assume that $\Delta$ has a
non-zero VEV, $\langle\Delta\rangle=V diag(1,-1)$ with $V\approx (O)$ TeV.
We can thus achieve the desired splitting in the MSSM Higgs doublets. Likewise,
we can explain just-so Higgs splitting in $SO(10)$ by using the
45-dimensional Higgs since it contains the $\Delta(1,1,3)$ of 4-2-2. We may
consider the non-renormalizable coupling $10\cdot10\cdot 45^2$
($10\cdot10\cdot 45$ is not allowed as $45$ is a two-index
anti-symmetric representation) in the superpotential,
where the $10$ is the 10-dimensional representation that has the MSSM Higgs doublets.
We may get just-so Higgs splitting by writing the corresponding SSB
term with a suitable choice of Yukawa coupling or VEV for the 45.

The fundamental parameters of
the 4-2-2 model that we consider are as follows:

\begin{align}
m_{0}, m_{H_u}, m_{H_d}, M_2, M_3, A_0, \tan\beta, {\rm sign}(\mu).
\label{params}
\end{align}
Here $m_0$ is the universal SSB mass for MSSM sfermions,
$A_0$ is the universal SSB trilinear scalar interaction (with the
corresponding Yukawa coupling factored out), $\tan\beta$ is the
ratio of the vacuum expectation values  (VEVs) of the two MSSM Higgs
doublets, and the magnitude of $\mu$, but not its sign, is determined by the
radiative electroweak breaking (REWSB) condition.
Although not required, we will
assume that the gauge coupling unification condition $g_3=g_1=g_2$
holds at $M_{\rm GUT}$ in 4-2-2. Such a scenario can arise,
for example, from a higher dimensional $SO(10)$
\cite{Hebecker:2001jb} or $SU(8)$ \cite{su8} model after suitable
compactification.

\section{Phenomenological constraints and scanning procedure\label{constraintsSection}}

We employ the ISAJET~7.80 package~\cite{ISAJET}  to perform random
scans over the parameter space listed in Eq.(\ref{params}). In this
package, the weak scale values of gauge and third generation Yukawa
couplings are evolved to $M_{\rm GUT}$ via the MSSM renormalization
group equations (RGEs) in the $\overline{DR}$ regularization scheme.
We do not strictly enforce the unification condition $g_3=g_1=g_2$ at $M_{\rm
GUT}$, since a few percent deviation from unification can be
assigned to unknown GUT-scale threshold
corrections~\cite{Hisano:1992jj}.
The difference between $g_1(=g_2)$ and $g_3$ at $M_{GUT}$ is no
worse than $4\%$. 
If neutrinos acquire mass via Type I seesaw, the impact of the neutrino Dirac
Yukawa coupling on the RGEs of the SSB terms, gauge couplings and the
third generation Yukawa
couplings is significant only for relatively large values ($\sim 2$ or so).
In the GYU 4-2-2 model we expect the largest (third family) 
Dirac Yukawa coupling to be
comparable to the gauge couplings ($\sim 0.6$ at $M_{\rm GUT}$).
Therefore, we do not include the Dirac neutrino Yukawa coupling
in the RGEs.

The various boundary conditions are imposed at
$M_{\rm GUT}$ and all the SSB
parameters, along with the gauge and Yukawa couplings, are evolved
back to the weak scale $M_{\rm Z}$.
In the evaluation of Yukawa couplings the SUSY threshold
corrections~\cite{Pierce:1996zz} are taken into account at the
common scale $M_{\rm SUSY}= \sqrt{m_{{\tilde t}_L}m_{{\tilde t}_R}}$. The entire
parameter set is iteratively run between $M_{\rm Z}$ and $M_{\rm
GUT}$ using the full 2-loop RGEs until a stable solution is
obtained. To better account for leading-log corrections, one-loop
step-beta functions are adopted for gauge and Yukawa couplings, and
the SSB parameters $m_i$ are extracted from RGEs at multiple scales
$m_i=m_i(m_i)$. The RGE-improved 1-loop effective potential is
minimized at an optimized scale $M_{\rm SUSY}$, which effectively
accounts for the leading 2-loop corrections. Full 1-loop radiative
corrections are incorporated for all sparticle masses.

The requirement of REWSB~\cite{Ibanez:1982fr} puts an important theoretical
constraint on the parameter space. Another important constraint
comes from limits on the cosmological abundance of stable charged
particles~\cite{Nakamura:2010zzi}. This excludes regions in the parameter space
where charged SUSY particles, such as ${\tilde \tau}_1$ or ${\tilde t}_1$, become
the lightest supersymmetric particle (LSP). We accept only those
solutions for which one of the neutralinos is the LSP and saturates
the WMAP (Wilkinson Microwave Anisotropy Probe) dark matter relic abundance bound.

We have performed random scans for the following parameter range:

\begin{align}0\leq  m_{0}, m_{H_u}, m_{H_d} \leq 20\, \rm{TeV} \nonumber \\
-2\rm{TeV} \leq M_2  \leq 2\,\rm {TeV} \nonumber \\
-2\rm{TeV} \leq M_3  \leq 2\, \rm{TeV} \nonumber \\
45\leq \tan\beta \leq 55 \nonumber \\
-3\leq A_{0}/m_0 \leq 3\nonumber \\
\mu < 0,\mu>0
 \label{parameterRange}
\end{align}
where $m_t = 173.3\pm1.1\, {\rm GeV}$ \cite{:1900yx} is the top quark pole mass. 
The value of the top quark mass is very crucial, as we shall see later, 
for GYU. 
We use $m_b(m_Z)=2.83$ GeV which is hard-coded into ISAJET. The above
choice of parameters is influenced by our previous experience with
the 4-2-2 model.

In scanning the parameter space, we employ the Metropolis-Hastings
algorithm as described in \cite{Belanger:2009ti}. All of the
collected data points satisfy
the requirement of REWSB,
with the neutralino in each case being the LSP.
We direct the Metropolis-Hastings algorithm
to search for solutions with GYU.
After collecting the data, we impose
the mass bounds on all the particles~\cite{Nakamura:2010zzi} and use the
IsaTools package~\cite{Baer:2002fv}
to implement the following phenomenological constraints on points that
have GYU to within 20\%:
\begin{table}[h]\centering
\begin{tabular}{rlc}
$m_h~{\rm (lightest~Higgs~mass)} $&$ \geq\, 114.4~{\rm GeV}$  
&  \cite{Schael:2006cr}                                                   \\
$BR(B_s \rightarrow \mu^+ \mu^-) $&$ <\, 5.8 \times 10^{-8}$                     
&   \cite{:2007kv}                                                        \\
$2.85 \times 10^{-4} \leq BR(b \rightarrow s \gamma) $ & 
$ \leq\, 4.24 \times 10^{-4} \; (2\sigma)$ &   \cite{Barberio:2008fa}     \\
$0.15 \leq \frac{BR(B_u\rightarrow \tau \nu_{\tau})_{\rm MSSM}}
{BR(B_u\rightarrow \tau \nu_{\tau})_{\rm SM}}$&$ \leq\, 2.41 \; (3\sigma)$ 
&   \cite{Barberio:2008fa}                                                 \\
$\Omega_{\rm CDM}h^2 $&$ =\, 0.111^{+0.028}_{-0.037} \;(5\sigma)$               
&  \cite{Komatsu:2008hk}                                                    \\
$0\leq \Delta (g-2)_{\mu}/2 $&$ \leq\, 55.6 \times 10^{-10}$   
&  \cite{Bennett:2006fi}
\end{tabular}
\end{table}

In the case of $\Delta (g-2)_{\mu}$, we only require that the
GYU 4-2-2 model does no worse than the SM. However,
we do give examples of solutions that satisfy the
$\Delta (g-2)_{\mu}/2$ constraint to within $3\sigma$.

\section{Threshold corrections and Yukawa unification\label{muneg}}

The SUSY threshold corrections to the top, bottom and tau Yukawa couplings
play a crucial role in $t$-$b$-$\tau$ Yukawa coupling unification.
In general, the bottom Yukawa coupling $y_b$ can receive large
threshold corrections, while the threshold corrections to $y_t$ are
typically smaller~\cite{Pierce:1996zz}. The scale at which Yukawa coupling unification 
occurs is set equal to $M_{\rm GUT}$, the scale of gauge coupling unification. 
Consider first the case $y_t ( M_{\rm GUT})\approx y_{\tau}(M_{\rm
GUT})$. The SUSY correction to the
tau lepton mass  is given by $\delta m_{\tau}=v\cos\beta \delta y_{\tau}$. 
For the large $\tan\beta$ values of interest here, there is sufficient 
freedom in the choice of $\delta y_{\tau}$ to achieve $y_t\approx y_{\tau}$ 
at $M_{\rm GUT}$. This freedom stems from the fact that 
$\cos\beta \simeq 1/\tan\beta$ for large $\tan\beta$, and so we may 
choose an appropriate $\delta y_{\tau}$ and $\tan\beta$ to give us both 
the correct $\tau$ lepton mass and $y_t \approx y_{\tau}$. 
The SUSY contribution to $\delta y_b$ has to be carefully monitored
in order to achieve Yukawa coupling unification $y_t (M_{\rm
GUT})\approx y_{b}(M_{\rm GUT}) \approx y_{\tau}(M_{\rm GUT})$.

We choose the sign of  $\delta y_i\, (i=t,b,\tau)$ from the perspective of
evolving $y_i$ from $M_{\rm GUT}$ to $M_{\rm Z}$. With this choice,
$\delta y_b$ must receive a negative contribution
($-0.27 \lesssim \delta y_b/y_b \lesssim -0.15$) in order to realize
Yukawa coupling unification~\cite{Gogoladze:2010fu}. This is a narrow interval considering
the full range of $-0.4\lesssim \delta y_b/y_b \lesssim 0.6$ that 
we found in the data that we collected.
The dominant contribution to $\delta y_b$ comes from
the finite corrections of the gluino and chargino loops,
and in our sign convention, is approximately given by~\cite{Pierce:1996zz}

\begin{align}
\delta y_b^{\rm finite}\approx\frac{g_3^2}{12\pi^2}\frac{\mu m_{\tilde g}
\tan\beta}{m_{\tilde b}^2}+
                         \frac{y_t^2}{32\pi^2}\frac{\mu A_t \tan\beta}{m_{\tilde t}^2},
\label{finiteCorrectionsEq}
\end{align}
where $g_3$ is the strong gauge coupling, $m_{\tilde g}$ is
the gluino mass, $m_{\tilde b}$ and $m_{\tilde t}$ are the lighter
sbottom and stop masses, and $A_t$ is the 
top trilinear (scalar) coupling.

The logarithmic corrections to $y_b$ are positive, which
leaves the finite corrections to provide for the correct overall negative
$\delta y_b$ in order to realize Yukawa unification. The gluino contribution 
(Eq.(\ref{finiteCorrectionsEq})) is positive for $\mu>0$ and same sign 
gaugino soft mass terms. Thus, the chargino 
contribution (Eq.(\ref{finiteCorrectionsEq})) must play an essential role
to provide the required negative contribution to $\delta y_b$. This can be
achieved with suitably large values of both $m_0$ and $A_t$.
This large value of $m_0$ and $A_t$ is the reason behind the requirement
of $m_0\gtrsim 6\,{\rm TeV}$ and $A_0/m_0\sim -2.6$ in the $SO(10)$ model
discussed in \cite{Baer:2008jn}. The parameter $\tan\beta$ also lies in a 
narrow range $48\lesssim\tan\beta\lesssim 52$. A similar 
trend was shown in \cite{Gogoladze:2009ug}
for the 4-2-2 model with same sign
gauginos and $\mu>0$. The latter model displays Yukawa coupling 
unification consistent with WMAP data via bino-gluino coannihilation.

\begin{figure}[b!]
\centering
\subfiguretopcaptrue
\subfigure[\hspace {1mm}  $SO(10)$]{
\includegraphics[width=7.2cm]{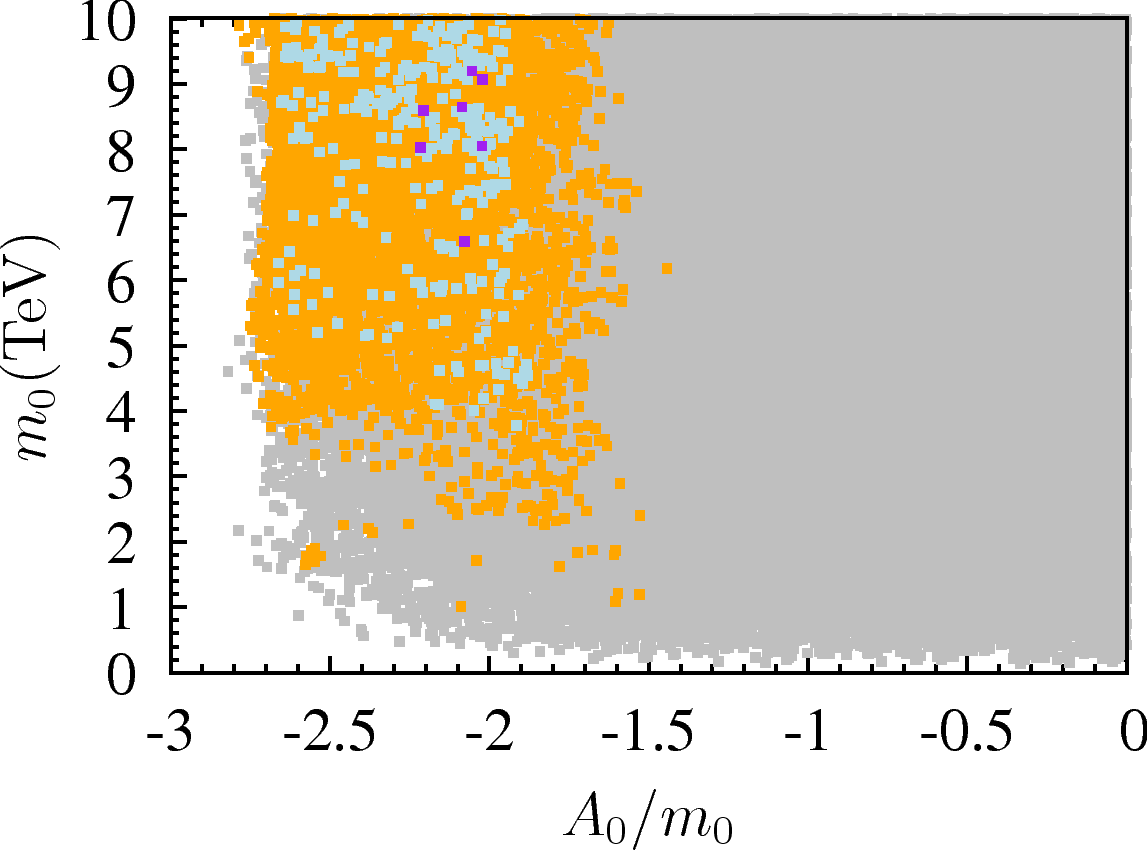}
}
\subfigure[\hspace {1mm}  $\mu>0,M_2>0,M_3>0$]{
\includegraphics[width=7.2cm]{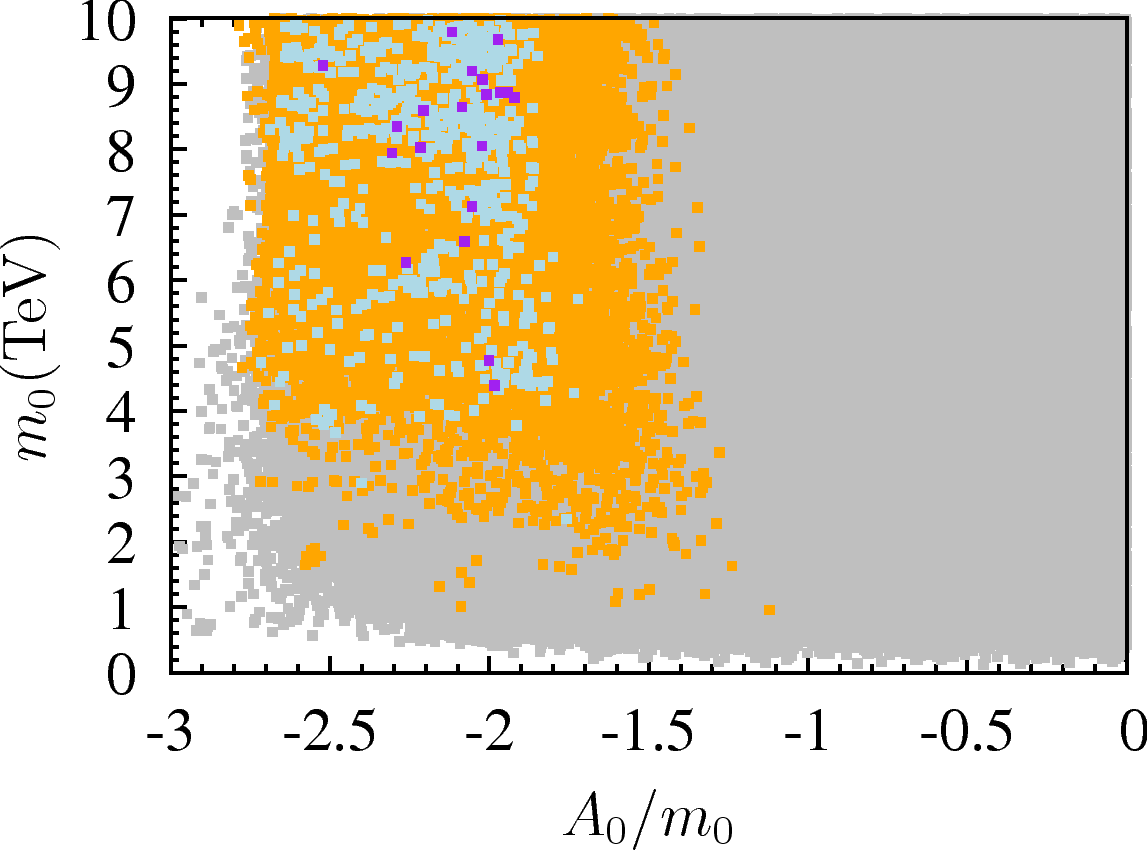}
}
\subfigure[\hspace {1mm}  $\mu<0,M_2<0,M_3>0$]{
\includegraphics[width=7.2cm]{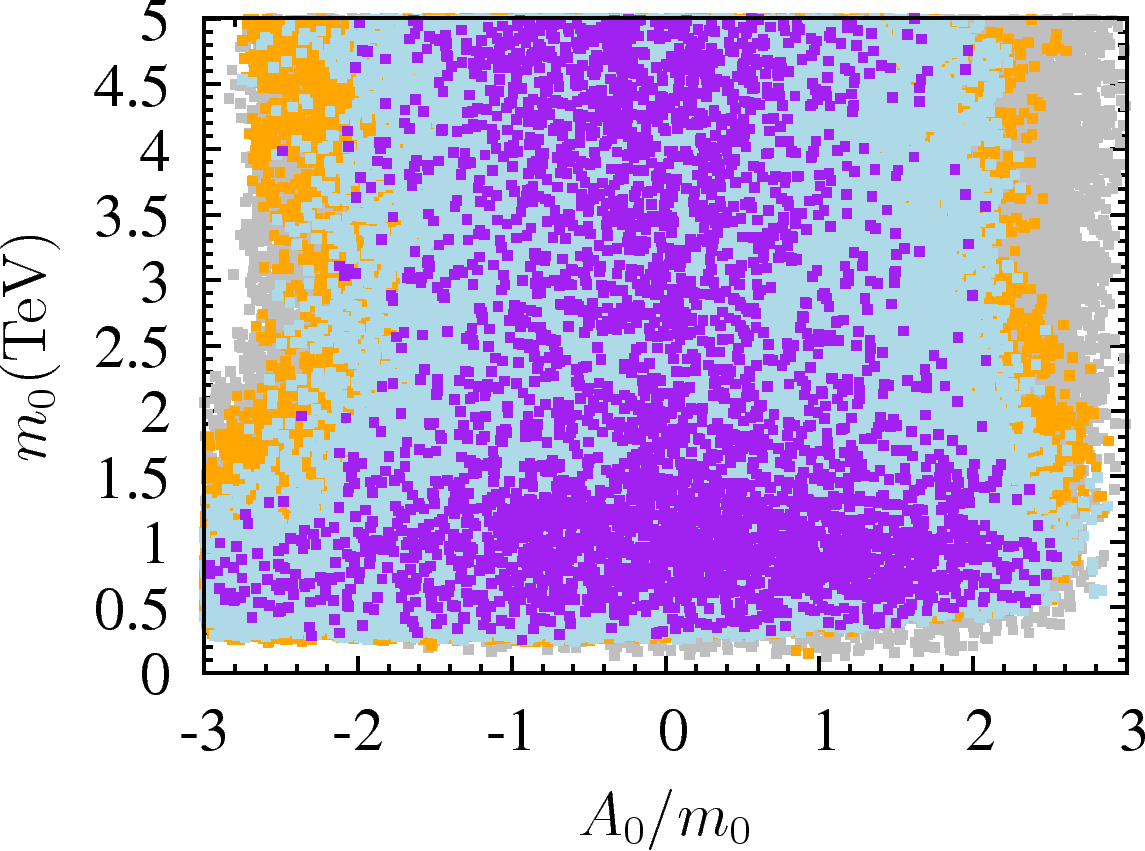}
}
\subfigure[\hspace {1mm}  $\mu>0,M_2>0,M_3<0$]{
\includegraphics[width=7.2cm]{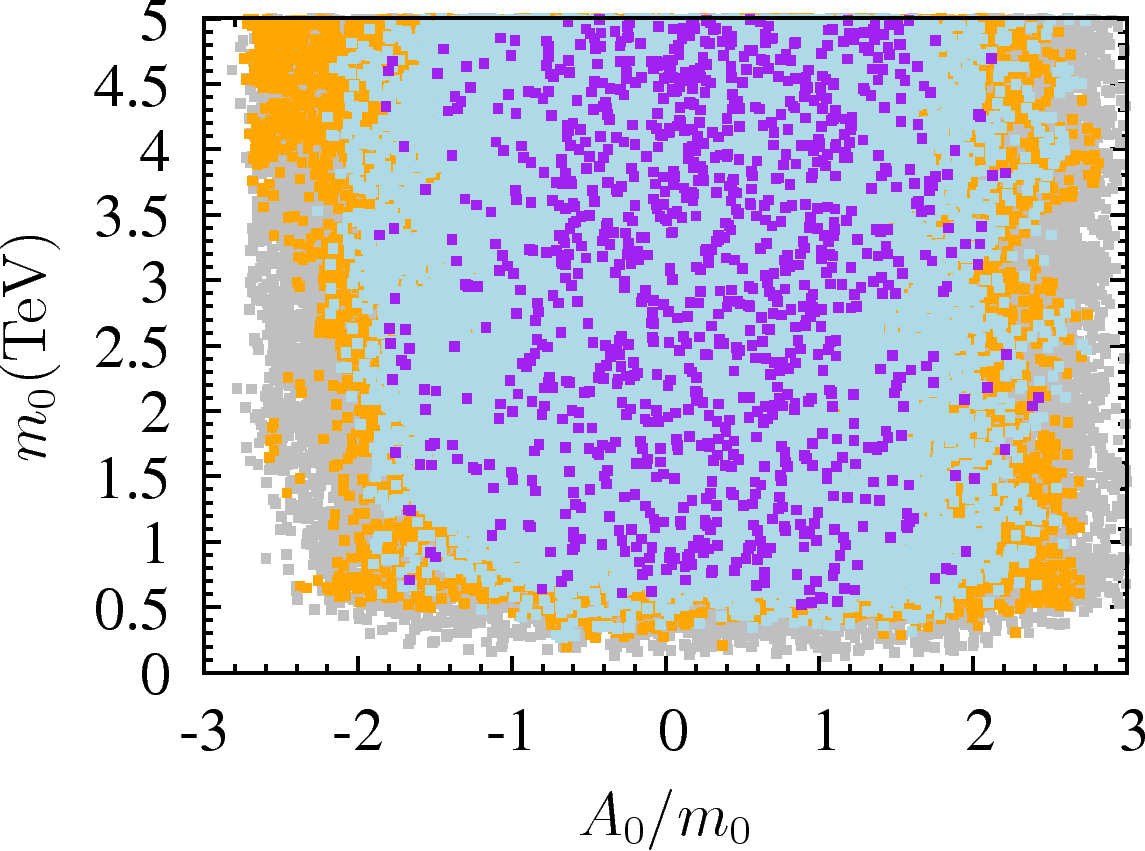}
}
\caption{Plots in the $m_0$ - $A_0/m_0$ plane for the $SO(10)$
model and three classes of 4-2-2 models. Gray points
are consistent with REWSB and $\tilde\chi_1^0$ LSP. Orange, light blue 
and purple points are subsets of gray points with $R\lesssim1.2,1.1,1.02$
respectively.}
\label{compareModels}
\end{figure}

In an $SO(10)$-like \cite{Baer:2008jn} model with same
sign gauginos, the case $\mu<0$ is not favored because of
the negative contribution to $\Delta(g-2)_\mu \propto \mu M_2$ which,
instead, needs to be positive. Therefore, while the Yukawa unified
$SO(10)$ and 4-2-2 \cite{Gogoladze:2009ug} (with $\mu,M_2,M_3>0$)
models do not provide the required contribution to $\Delta(g-2)_\mu$
because of heavy sparticles, they do no worse than the SM in
this respect.

One can improve the situation immensely by considering the case
of opposite sign gaugino soft terms which is allowed by the 4-2-2
model. We showed in \cite{Gogoladze:2010fu} the parameter space
corresponding to $\mu,M_2<0,M_3>0$ that gives Yukawa
coupling unification with a sub-TeV sparticle
spectrum which is consistent with all known experimental bounds
including $\Delta(g-2)_\mu$. Another possibility is to
consider $\mu,M_2>0,M_3<0$ (the parameter space
for this case has not been previously discussed in the literature).
This becomes possible because the gluino contribution to
$\delta y_b$ is of the correct (negative) sign.

In order to quantify Yukawa coupling unification, following
\cite{Baer:2008jn}, we define the quantity $R$ as,
\begin{align}
R=\frac{ {\rm max}(y_t,y_b,y_{\tau})} { {\rm min} (y_t,y_b,y_{\tau})}
\end{align}
Thus, $R$ is a useful indicator for Yukawa unification with $R\lesssim 1.1$,
for instance, corresponding to Yukawa unification to within 10\%, while
$R=1.0$ denotes `perfect' Yukawa unification.

In Figure~\ref{compareModels} we show a comparison between the four
model types considered, {\it i.e.} the $SO(10)$ model and 4-2-2 models
with $\{\mu>0,M_2>0,M_3>0\}$, $\{\mu<0,M_2<0,M_3>0\}$ and $\{\mu>0,M_2>0,M_3<0\}$.
We show plots in the $m_0$ - $A_0/m_0$ plane for these models. Gray points
are consistent with REWSB and $\chi_1^0$ LSP. Orange, light blue and purple
points are subsets of gray points with $R\lesssim1.2,\,1.1,\,1.02$
respectively. As previously explained, in the $SO(10)$ and 
4-2-2 models with $\mu>0$ and same sign gauginos, Yukawa coupling
unification can only be achieved for large values of $m_0$. Also, the value
of $A_0/m_0$ is very restricted. On the other hand,
with opposite sign gauginos one can realize very credible Yukawa unification
with relatively small $m_0$ values. Also as previously described, since the gluino loop
provides the required $\delta y_b$, $A_0/m_0$ is
no longer restricted by Yukawa coupling unification and can vary over
a very wide range.

\begin{figure}[b!]
\centering
\subfiguretopcaptrue
\subfigure[\hspace {1mm}  Same Sign Gauginos]{
\includegraphics[width=7.2cm]{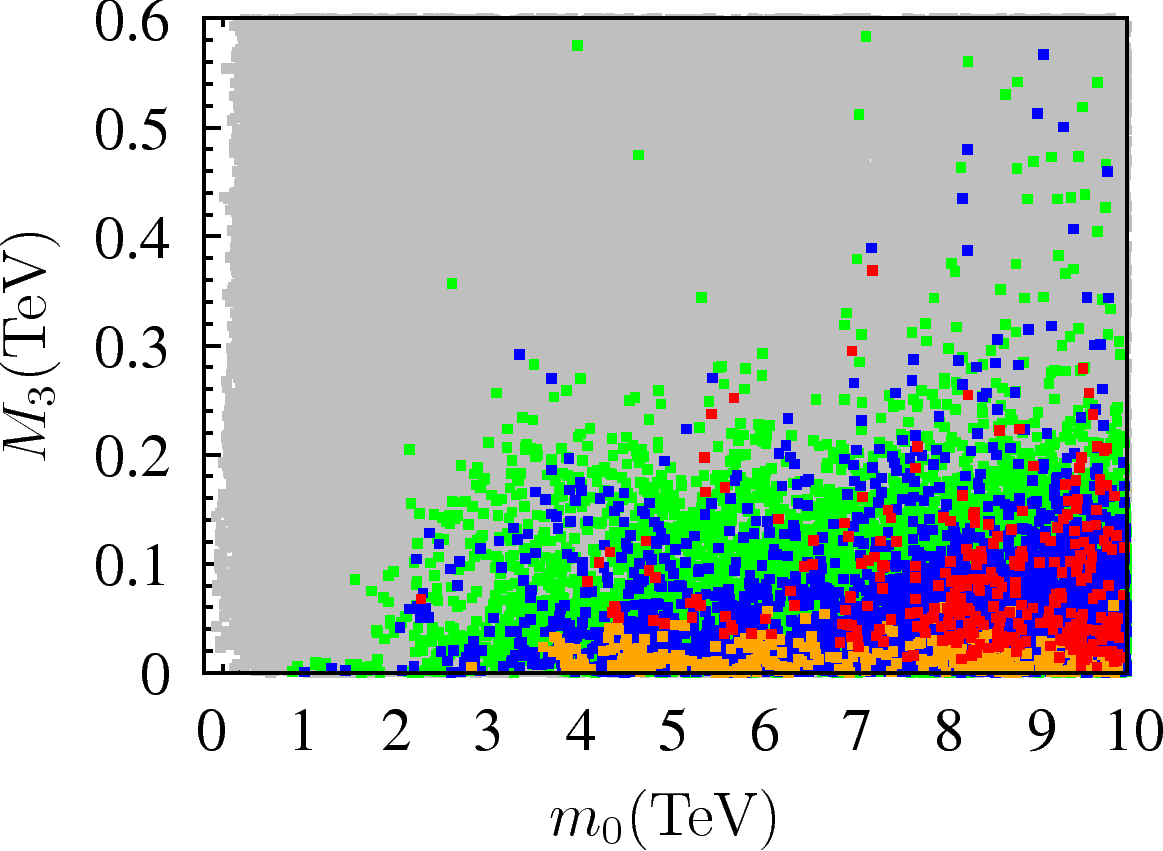}
}
\subfigure[\hspace {1mm}  Opposite Sign Gauginos]{
\includegraphics[width=7.2cm]{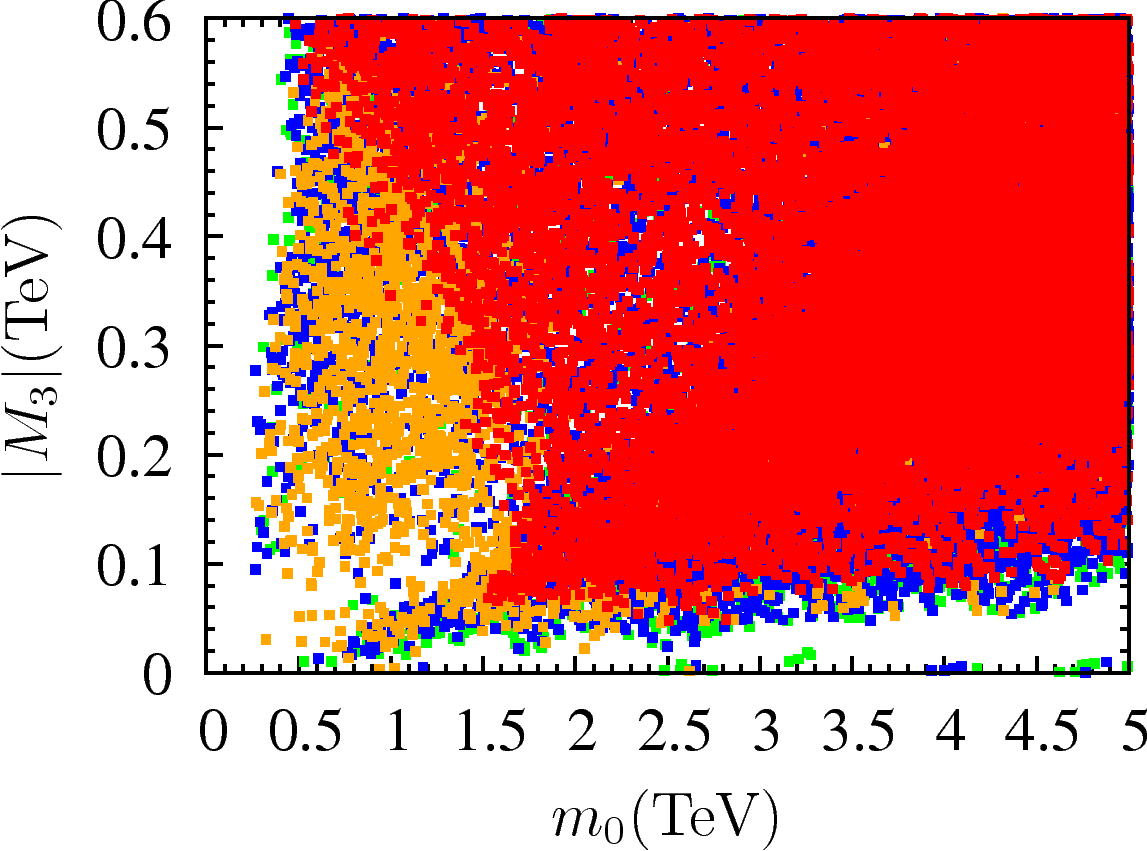}
}
\caption{Plots in the $\lvert M_3\rvert$ - $m_0$ plane for 
same sign (left) and opposite sign (right) gaugino (including points for both 
$\{\mu<0,M_2<0,M_3>0\}$ and $\{\mu>0,M_2>0,M_3<0\}$) cases. 
Gray points are consistent with REWSB and $\tilde\chi_1^0$ LSP. 
Green, blue 
and orange points are subsets of gray points with $R\lesssim1.2,1.15,1.1$
respectively. Red points satisfy particle mass bounds in addition
to $R\lesssim 1.1$.}
\label{compareModelsM3m0}
\end{figure}

\begin{figure}[t]
\centering
\includegraphics[width=7.2cm]{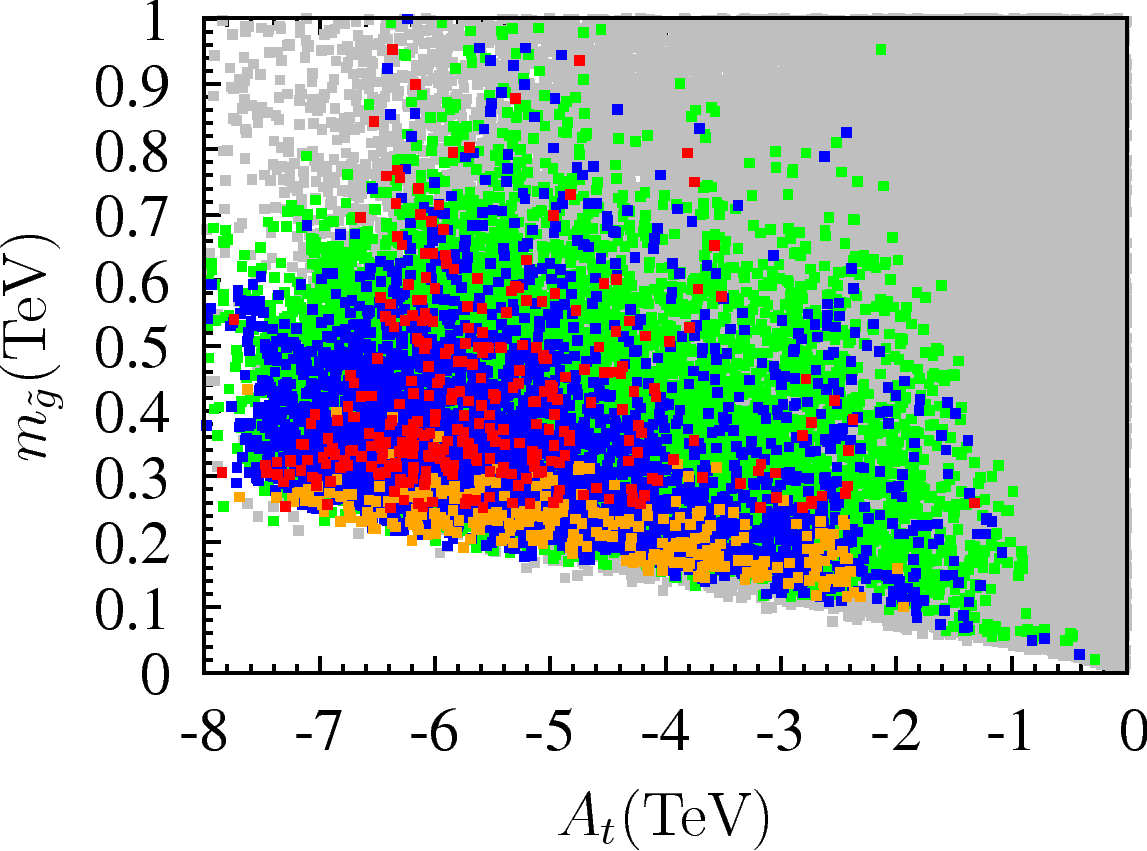}
\caption{Plot in the $m_{\tilde g}$ - $A_t$ plane plane for
same sign gauginos. Color coding same as in Figure~\ref{compareModelsM3m0}.
}
\label{sameSignmgAt}
\end{figure}

In Figure~\ref{compareModelsM3m0} we display an interesting 
difference between the same sign and opposite sign
gaugino cases in the  $\lvert M_3\rvert$ - $m_0$ 
plane. Shown in gray are points that satisfy the requirements of 
REWSB and $\tilde\chi_1^0$ LSP. In green, blue and orange, we show points that 
further satisfy Yukawa unification to within 20\%,15\% and 10\% respectively. 
Red points satisfy the particle mass bounds in addition to having $R\lesssim 1.1$.
The trend of a lower $M_3$ in the case of same sign gauginos is very apparent 
if we require Yukawa coupling unification. The reason for this is again that we 
need to suppress the finite correction to the bottom quark mass coming from the 
gluino loop (see Eq.(\ref{finiteCorrectionsEq})). The case of opposite sign gauginos, 
in stark contrast, shows that essentially any value of $\lvert M_3\rvert$ 
is acceptable as far as Yukawa coupling unification is concerned. The orange 
region in the bottom left in this case is excluded because 
of the lower bound on the gluino mass. It is also instructive to consider the 
$m_{\tilde g}$ - $A_t$ plane in the case of same sign gauginos shown in 
Figure~\ref{sameSignmgAt}. In this figure, $m_{\tilde g}$ is the physical gluino 
mass and $A_t$ is the value of the top trilinear (scalar) coupling at the scale 
$Q=\sqrt{m_{\tilde t_L}m_{\tilde t_R}}$. The color coding is the same 
as in Figure~\ref{compareModelsM3m0}. This figure shows that a lighter 
gluino is required for Yukawa coupling unification with a smaller absolute value 
of $A_t$. This again reflects the fact that for Yukawa coupling unification, 
we need to suppress the gluino contribution to $\delta y_b$ in favor of the 
chargino contribution.

Finally, it is interesting to note that with opposite sign gauginos, the MSSM 
parameter $\tan\beta$ varies over a wider range, 
$44\lesssim\tan\beta \lesssim 54$.
With same sign gauginos, on the other hand,  $48\lesssim\tan\beta \lesssim 52$. 
Among other interesting features of a Yukawa unified
model with opposite sign gauginos is the presence of various channels for realizing 
the desired $\tilde\chi_1^0$ relic density. In particular, in \cite{Gogoladze:2010fu}
we showed the existence of stau coannihilation, bino-wino coannihilation, gluino 
coannihilation and CP-odd Higgs resonance solutions for the case $(\mu<0,M_2<0,M_3>0)$. 
In contrast, for the Yukawa unified $SO(10)$ model, only 
the light Higgs resonance solution is consistent with the WMAP
relic density. It is interesting to note that in the case of
4-2-2 models (with $\mu>0, M_2>0, M_3>0$),
it is not possible to get the well-known stau ($\tilde \tau$) coannihilation channel.
This is because in the $\tilde \tau$ mass$^2$ matrix the diagonal terms 
are proportional to $m_0^2$, whereas the off-diagonal terms
are proportional to $A_\tau m_\tau$, where 
$A_\tau$ is the low-scale value of the tau trilinear (scalar) coupling. For
$\mu>0, M_2>0, M_3>0$, one needs a heavy $m_0$ in order to
realize Yukawa coupling unification. One therefore needs $A_\tau\sim m_0^2/m_\tau$
in order for the off-diagonal terms to contribute to give a small
stau mass $m_{\tilde\tau} \sim m_{\tilde\chi_1^0}$, where
$\tilde\chi_1^0$ is the lightest neutralino. This, clearly, is not possible
for large $m_0$ values. The parameter space of the $SO(10)$ model is a
subset of the 4-2-2 model with $\{\mu>0, M_2>0, M_3>0\}$
and so these remarks apply to $SO(10)$ as well.

\section{Gauge-Yukawa unification \label{gaugeYukawaSec}}

In this section, we first describe a specific model where 
GYU may happen. We then move on to discuss SUSY 
particle thresholds and their effects on analyzing
GYU. 
It is helpful to define, in analogy with $R$, a parameter
$GY$ that quantifies GYU;
\begin{align}
GY=\frac{{\rm max}(g_1,g_2,g_3,y_t,y_b,y_{\tau})}{{\rm min}(g_1,g_2,g_3,y_t,y_b,y_{\tau})}
\end{align}

\subsection{Model for Gauge-Yukawa unification}

A six dimensional model realizing unification of the gauge couplings ($g_1$,
$g_2$, $g_3$) and the third family Yukawa couplings ($y_t$, $y_b$,
$y_{\tau}$) was presented in \cite{su8}. It has 
$SU(8)$ gauge symmetry with N=2 SUSY, which corresponds
to N=4 SUSY in 4D, and thus only the gauge multiplet can be introduced
in the bulk. The 6D N=2 gauge multiplet, expressed in terms of 4D,
N=4 gauge multiplet, contains the vector multiplet
$V(A_{\mu}, \lambda)$ and three chiral multiplets in the adjoint
(63-dimensional) representation of the gauge group.  The
63-dimensional gauge  multiplet contains the gauge bosons (and
their superpartners), while the three 63-dimensional chiral
multiplets contain the third family matter fermions and the Higgs
bosons plus their superpartners. The two extra dimensions are
compactified on the orbifold $T^2/Z_6$, and a suitable choice of the
$Z_6$ transformation matrix breaks $SU(8)$ down to $SU(4)
\times SU(2)_L\times SU(2)_R\times U(1)^2$. The theory reduces
to 4D N=1 SUSY 4-2-2 model and two additional $U(1)$ factors. 
The massless modes after compactification are the
4-2-2 gauge fields, $\mathbf{(15, 1, 1), (1, 3, 1), (1, 1, 3)}$
two singlet vector fields $\mathbf{(1, 1, 1)}$
and $\mathbf{(1, 1, 1)}$, third-family
matter fermions $\Psi_L=\mathbf{(4,\, 2,\, 1)}_{2,\,0}$ and
$\Psi_{\bar R}=\mathbf{(\bar{4},\,1,\, 2)}_{-2,\, -4}$, and the
bi-doublet Higgs fields, $H_1=\mathbf{(1,\, 2, \,2)}_{0,\, 4}$
and $H_2=\mathbf{(1,\, 2, \, 2)}_{0,\,-4}$. 

The trilinear coupling for the chiral multiplets
\begin{equation}\label{aa1}
S=\int d^6 x \left[\int d^2 \theta \,
2\, 
{\rm
Tr}\left(-\sqrt{2} g_6 \Sigma [\Phi, \Phi ^c]\right)+h.c.\right]
\end{equation}
includes the third family Yukawa interaction terms
\begin{equation}\label{aa2}
S=\int d^6 x \int d^2 \theta \,y_6 {\Psi}_L H_1 \Psi_{\bar R}
+h.c.
\end{equation}
In Eq. (\ref{aa1}), $\Sigma, \, \Phi,\, \Phi^c$  are chiral
multiplets containing the third family chiral fields, $\Psi_L$
and $\Psi_{\bar R}$, and the bi-doublet Higgs fields, $H_1$
and $H_2$, and $g_6$ and $y_6$ are the 6D gauge
and Yukawa couplings. Eqs. (\ref{aa1}) and (\ref{aa2}) lead to
$g_6=y_6$ with proper renormalization of the kinetic terms. Integrating out
the two extra dimensions, we obtain $y_4=g_4$ for the 4D
coupling leading to
\begin{equation}\label{aa3}
g_1=g_2=g_3=y_t=y_b=y_{\tau}(=y_{\nu_\tau}^{\rm Dirac})
\end{equation}
at the compactification scale ($M_c$) which we identify with the 
four dimensional unification scale $M_{\rm GUT}$. 
We assume that the 4-2-2 symmetry, as well as the two extra $U(1)$ 
are broken at $M_{\rm GUT}$
to the $SU(3)_c\times SU(2)_L \times U(1)_Y$ using suitable
Higgs vevs on the brane. We further assume that the breaking of the 
two extra $U(1)$ symmetries does not cause $D$ term splittings so 
that the universality of squark mass$^2$ terms at $M_{\rm GUT}$ 
is preserved. The first and second families are 
treated as brane fields to cancel the brane localized gauge anomalies. 
The Yukawa couplings for the first and second families are suppressed by a 
large volume factor, but there is no good reason as to why the 
mass of the first family is hierarchically small. The particle spectrum
below $M_{\rm GUT}$ is the same as in MSSM. 
For related discussions see Ref. \cite{Burdman:2002se}.

\subsection{SUSY thresholds and Gauge-Yukawa unification}

\begin{figure}[b!]
\centering
\includegraphics[width=7.2cm]{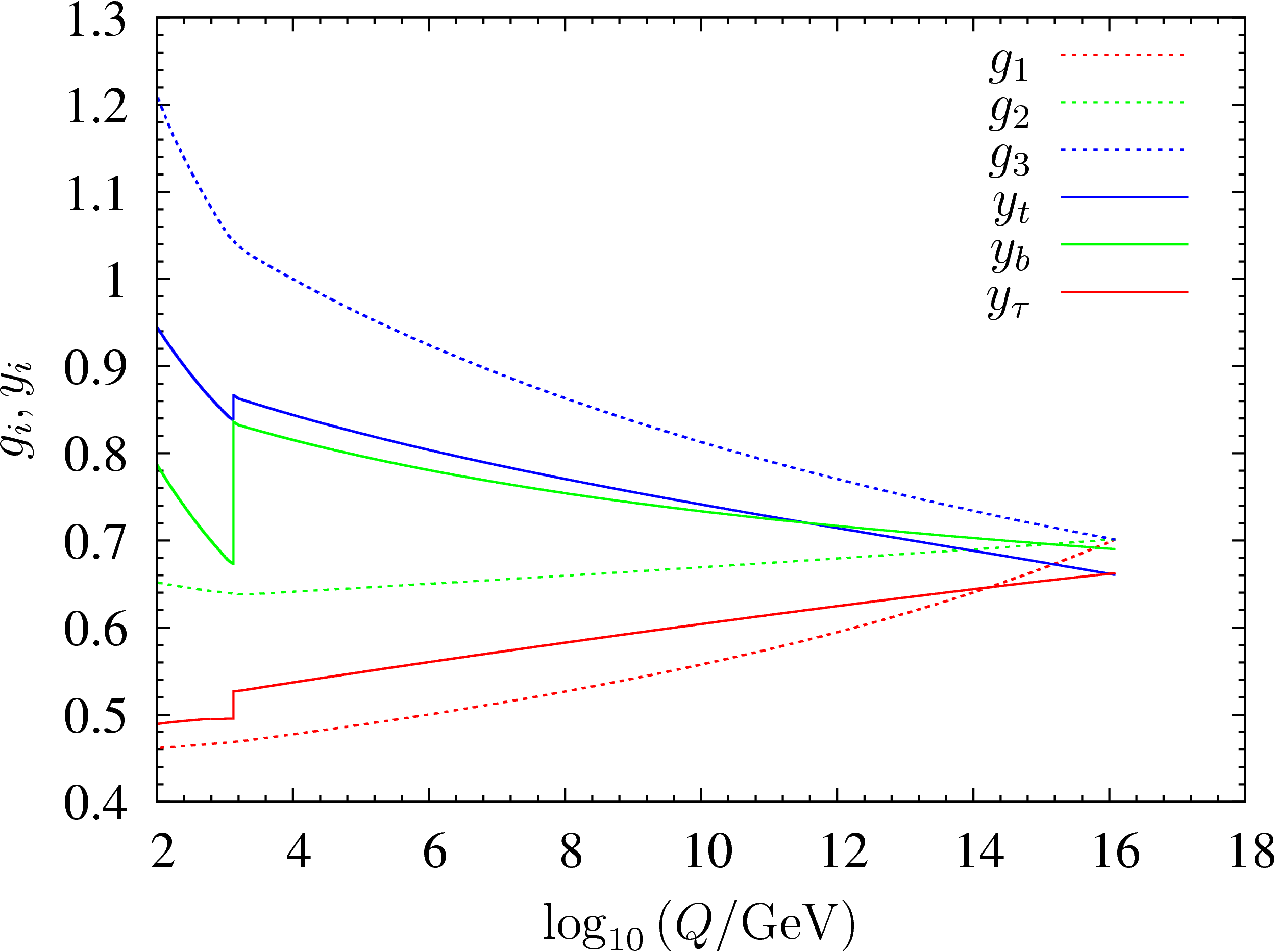}
\caption{ Gauge-Yukawa unification to within 6\% in
$SU(4)_c\times SU(2)_L\times SU(2)_R$ (4-2-2).
\label{runningFig}}
\end{figure}

As previously mentioned, the bottom quark and tau lepton Yukawa couplings receive 
larger threshold corrections than $y_t$. 
Since the gauge coupling is more or less fixed $\sim 0.69$, and since
Yukawa coupling unification typically occurs for
$y_t\approx y_b\approx y_{\tau}\sim 0.6$, the quantity $GY\sim 1.15$ for
the Yukawa unified models discussed above. If we desire to impose
GYU on our models, $y_t$ becomes the bottleneck for
a given top mass as typically $y_b$ and $y_{\tau}$ can be made larger than
$y_t$ by a suitable choice of the SUSY spectrum. 
In particular, the values of $y_b$ and $y_{\tau}$ at $M_{\rm GUT}$ can
be pushed up to $y_b(M_{\rm GUT})\sim 1.2 g_{\rm GUT}$ and
$y_{\tau}(M_{\rm GUT})\sim 1.2 g_{\rm GUT}$, where $g_{\rm GUT}$ is the
value of the unified gauge coupling at $M_{\rm GUT}$. The
leading SUSY threshold correction to the top quark mass is given by 
\cite{Pierce:1996zz}
\begin{align}
\delta y_t^{\rm finite}\approx\frac{g_3^2}{12\pi^2}\frac{\mu m_{\tilde g}
\tan\beta}{m_{\tilde t}^2}
\label{topThresh}
\end{align}

In our sign convention (evolving the couplings from $M_{\rm GUT}$ to $M_Z$),
a negative contribution to $\delta y_t$ is preferred.
Naively, a larger negative contribution allows for a 
larger $y_{t} (M_{\rm GUT})$. However, in the case of same sign
gauginos with $\mu>0$, we get a positive contribution to $\delta y_t$,
in which case a large $m_0$ value is required.
The requirement of a large $m_0$ can be argued from just
requiring Yukawa coupling unification. The significance of looking
at the sign of the correction to $\delta y_t$ in this case is the
realization that it may not be possible to achieve (more or less) gauge-Yukawa
unification at all. We see that $GY\gtrsim 1.13$ in the data 
that we have collected. In the case of opposite sign gauginos, on 
the other hand, our choice of the sign of $\mu$ gives a negative 
contribution to $\delta y_t$. We should, therefore, 
expect that GYU is allowed in the case 
of opposite sign gauginos.

\begin{figure}[b!]
\centering
\includegraphics[width=7.2cm]{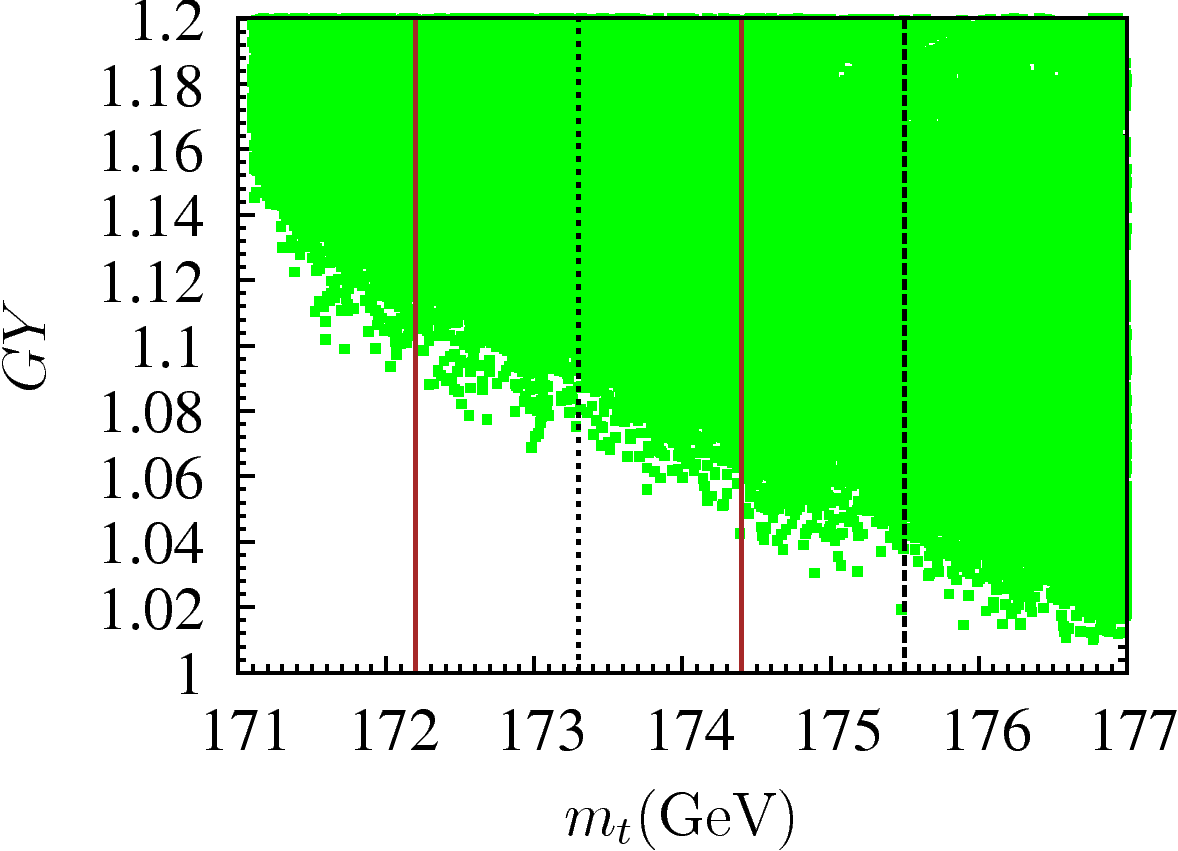}
\caption{Plot of $GY$ versus $m_t$. The vertical lines
correspond to $m_t=$ 172.2, 173.3, 174.4 and 175.5 GeV.
\label{topVaryFig}}
\end{figure}

In Figure~\ref{runningFig}  
we show the evolution of the gauge couplings and the third generation
Yukawa couplings that unify to within 6\% in the 4-2-2 model. The spectrum for this
point is given as Point $4$ in Table~\ref{table1}.

\boldmath\subsection{Gauge-Yukawa unification and ${m_t}$ \label{topVarySec}}\unboldmath

It is perhaps not too surprising that the parameter $GY$, a measure of 
GYU, depends sensitively on the top quark mass $m_t$.
It is, therefore,
instructive to study how GYU is affected as one 
varies $m_t$. We plot in Figure~\ref{topVaryFig} $GY$ as a
function of $m_t$. As expected, GYU prefers a 
larger top mass, with near perfect unification possible for $m_t=177\,{\rm GeV}$.
We next discuss GYU
allowing for a $1\sigma$ variation in the top mass.

\section{Gauge-Yukawa unification and sparticle spectro-\\
scopy \label{results}}

\begin{figure}[b!]
\centering
\includegraphics[width=8.4cm]{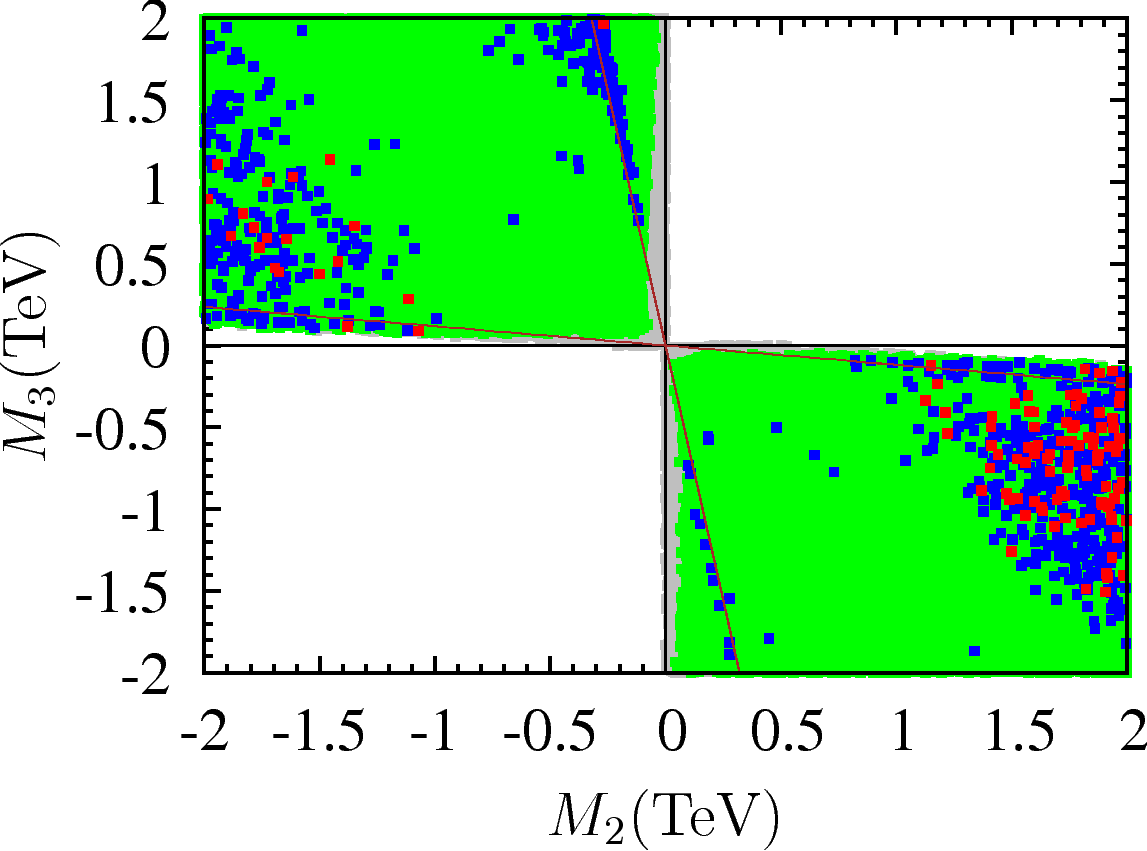}
\caption{
Results in the $M_3$ - $M_2$ plane. Gray points are consistent
with REWSB and $\tilde{\chi}^0_{1}$ LSP.
Green points satisfy particle mass bounds and constraints from
$BR(B_s\rightarrow \mu^+ \mu^-)$, $BR(b\rightarrow s \gamma)$
and $BR(B_u\rightarrow \tau \nu_\tau)$. In addition, we require that
green points do no worse than the SM in terms of $(g-2)_\mu$.
Blue points belong to a subset of green points and satisfy
the WMAP bounds on $\tilde{\chi}^0_1$ dark matter abundance.
Points in red represent the subset of blue
points satisfying gauge-Yukawa coupling unification to within 10\%.
We also show the lines $2 M_3=-13 M_2$ and $41 M_3=-6 M_2$ discussed 
in the text. 
}
\label{M3M2Fig}
\end{figure}

\begin{figure}[t!]
\centering
\includegraphics[width=7.2cm]{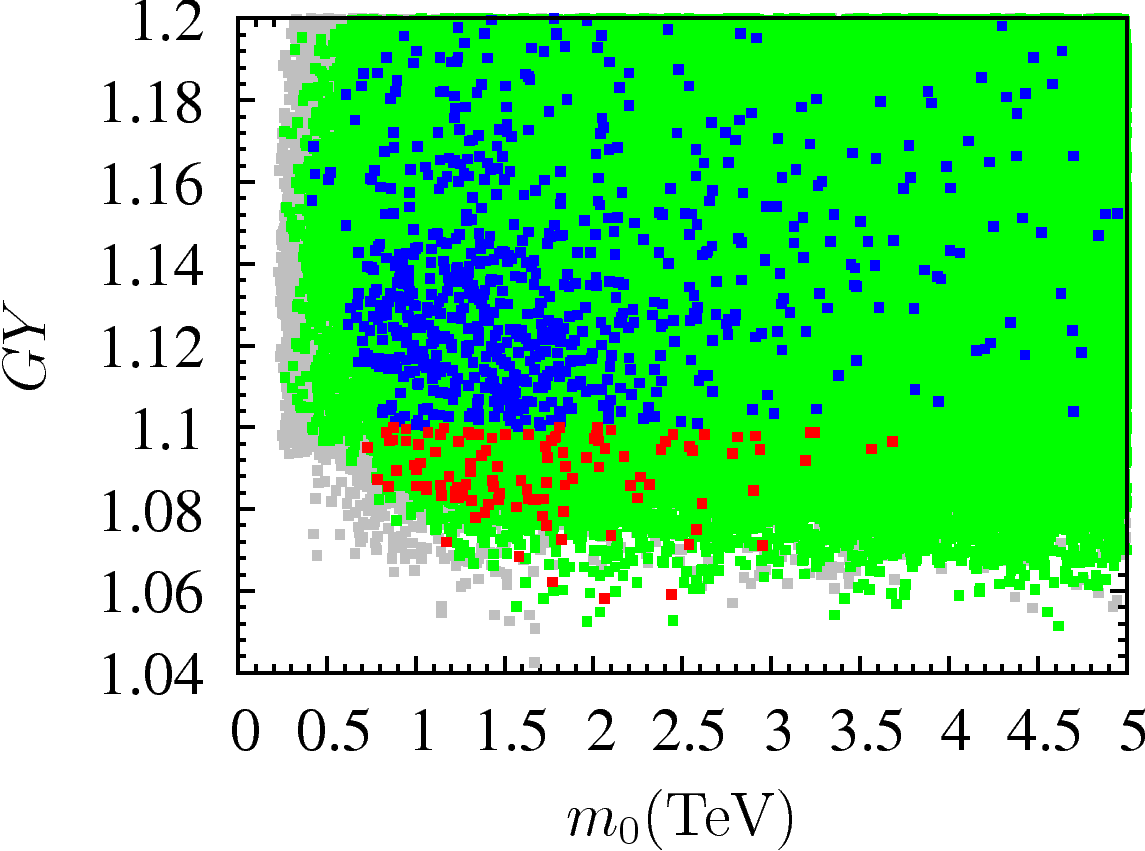}
\includegraphics[width=7.2cm]{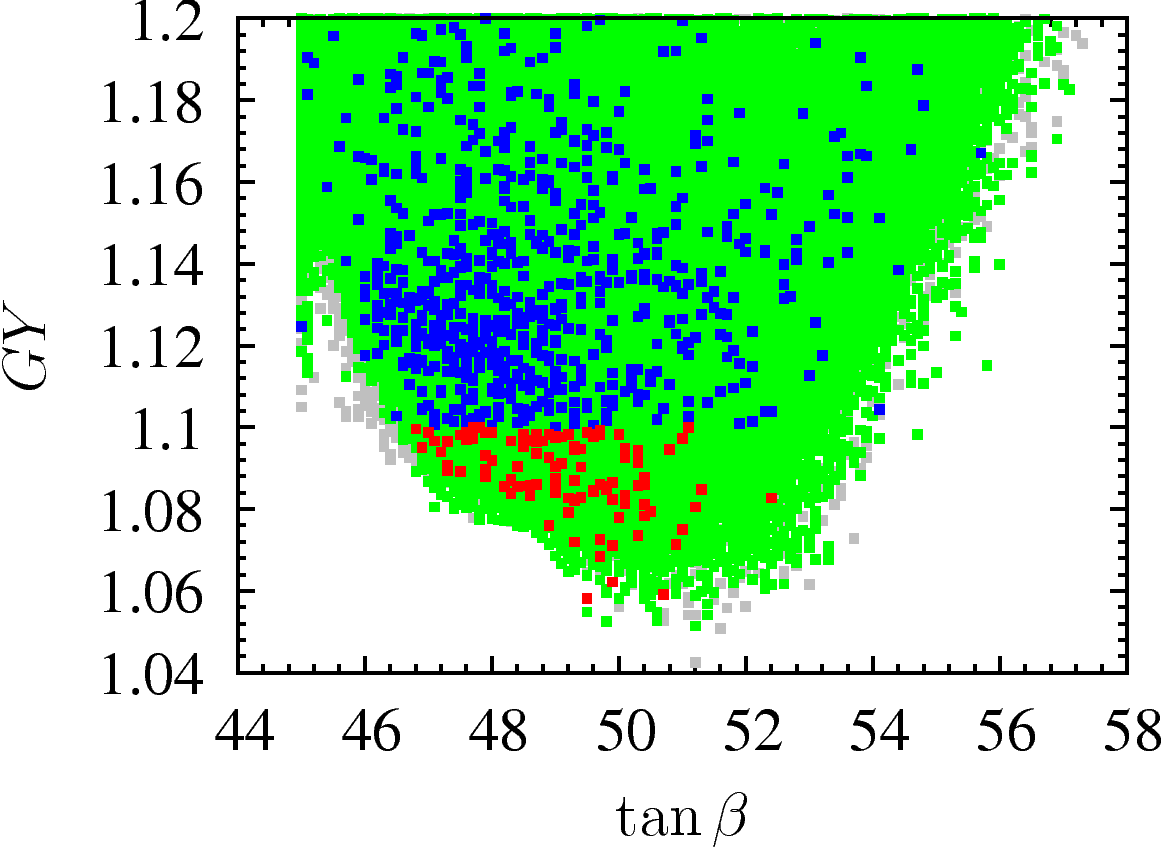}
\caption{Plots in the $GY$ - $m_0$ and $GY$ - $\tan\beta$ planes.
The two classes of opposite sign gaugino models are shown together.
Color coding same as in Figure~\ref{M3M2Fig}.}
\label{GYm0tanbFig}
\end{figure}

\begin{figure}[b!]
\centering
\subfiguretopcaptrue
\subfigure[\hspace {1mm}  $\mu<0,M_2<0,M_3>0$]{
\includegraphics[width=7.2cm]{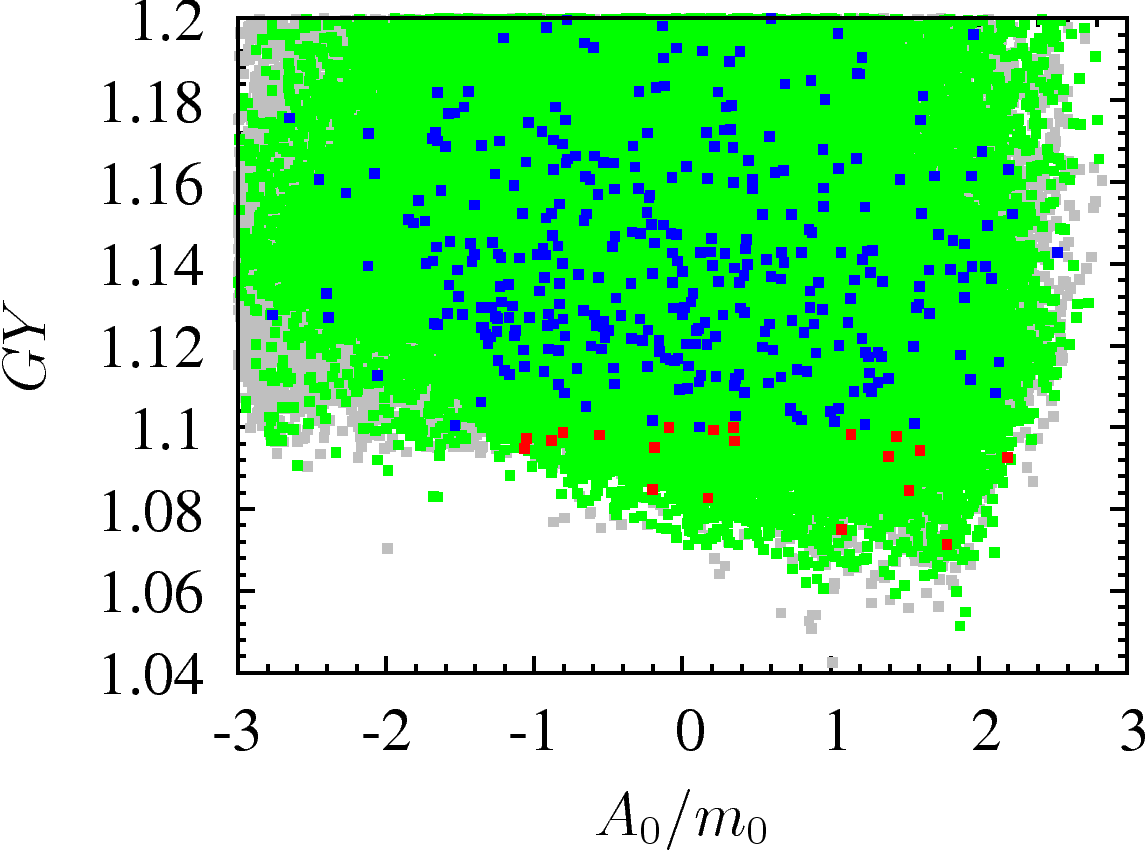}
}
\subfigure[\hspace {1mm}  $\mu>0,M_2>0,M_3<0$]{
\includegraphics[width=7.2cm]{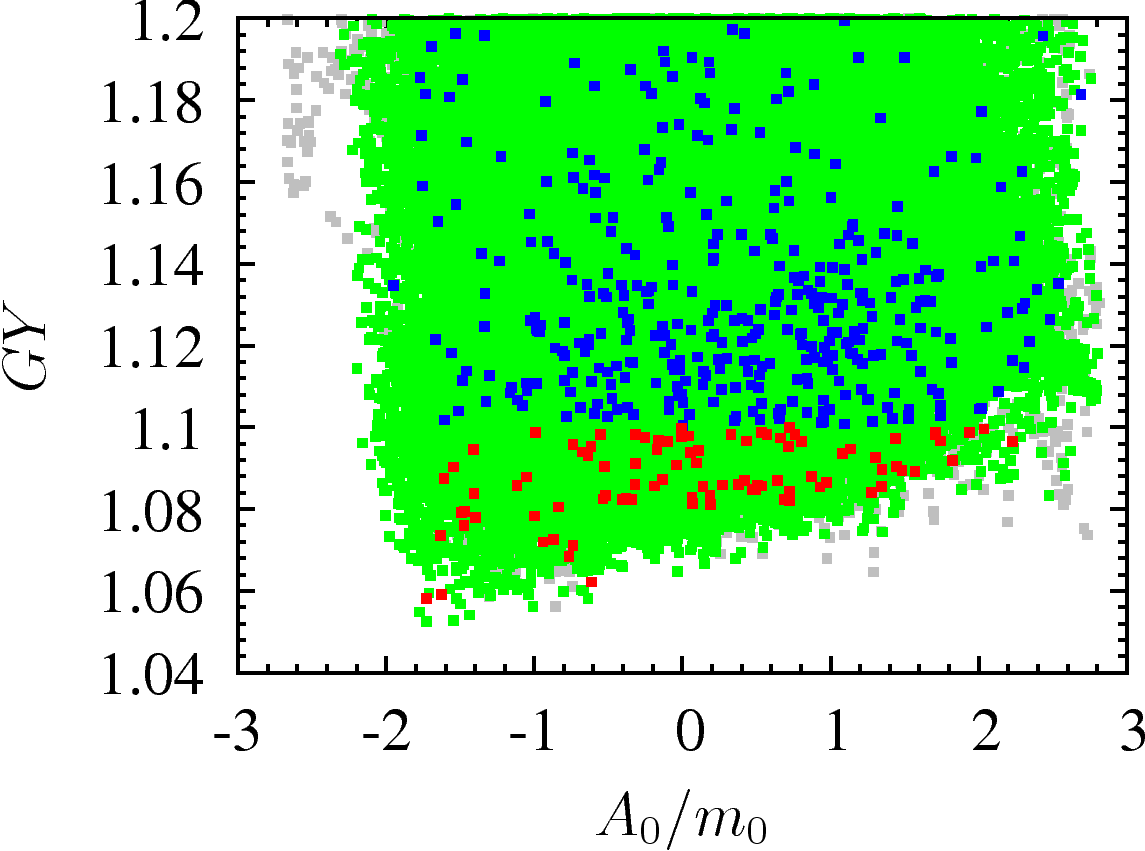}
}
\caption{Plots in the $GY$ - $A_0/m_0$ plane for
the two classes of opposite sign gaugino models.
Color coding same as in Figure~\ref{M3M2Fig}.}
\label{GYa0m0Fig}
\end{figure}

We present here the results of the scan over the parameter space
listed in Eq.(\ref{parameterRange}) after allowing for $m_t$ to vary
within $1\sigma$ of its central value. In Figure~\ref{M3M2Fig} we show
results in the $M_3$ - $M_2$ plane.
As previously explained,
gauge-Yukawa
unification prefers opposite sign gauginos. We emphasize this by only showing the 
two cases of relative sign gauginos and leaving two 
quadrants empty in Figure~\ref{M3M2Fig}. 
(Same sign gauginos GY unification 
of order 10\% or higher. See later.)
The gray points are consistent
with REWSB and $\tilde{\chi}^0_{1}$ LSP, while the 
green points also satisfy the particle mass bounds and constraints 
from $BR(B_s\rightarrow \mu^+ \mu^-)$, $BR(b\rightarrow s \gamma)$
and $BR(B_u\rightarrow \tau \nu_\tau)$. In addition, we require that the
green points fare no worse than the SM as far as $(g-2)_\mu$ is concerned.
The blue points belong to the subset of green points that satisfies
the WMAP bounds on $\tilde{\chi}^0_1$ dark matter abundance.
Points in red represent the subset of blue
points that satisfies gauge-Yukawa coupling unification to within 10\%.
We also show the lines $ M_3=-6.3 M_2$ and $M_3=-0.12 M_2$.
The slopes of these lines indicate bino-wino
and bino-gluino coannihilation in the $M_3$ - $M_2$ plane.
If we start off with a universal gaugino mass at $M_{\rm GUT}$,
we get $M_2/M_1(Q) \approx \pm1.89$ and $M_3/M_1 (Q) \approx \pm 4.67$, where the
negative sign is for the case of opposite sign gauginos. Therefore,
in order to get bino-wino coannihilation we should set 
$M_1/M_2 (M_{\rm GUT})\approx \pm 1.89$. Substituting this ratio of
$M_1$ and $M_2$ in Eq.~(\ref{gauginoCondition}) we can infer that
bino-wino coannihilation will occur for $ M_3\approx-6.3 M_2$. A similar
calculation shows that for bino-gluino coannihilation we should set 
$M_3\approx-0.12 M_2$.

\begin{figure}[b!]
\centering
\subfiguretopcaptrue
\subfigure{
\includegraphics[width=7.2cm]{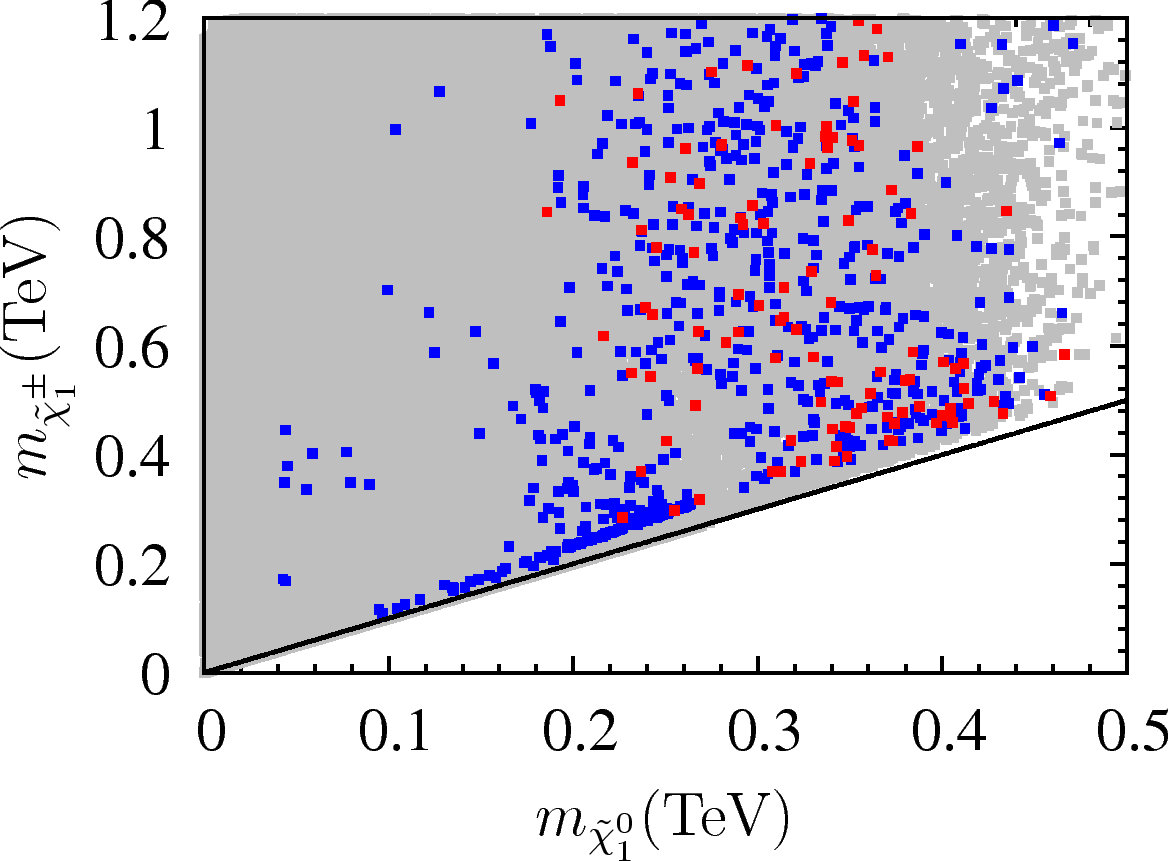}
}
\subfigure{
\includegraphics[width=7.2cm]{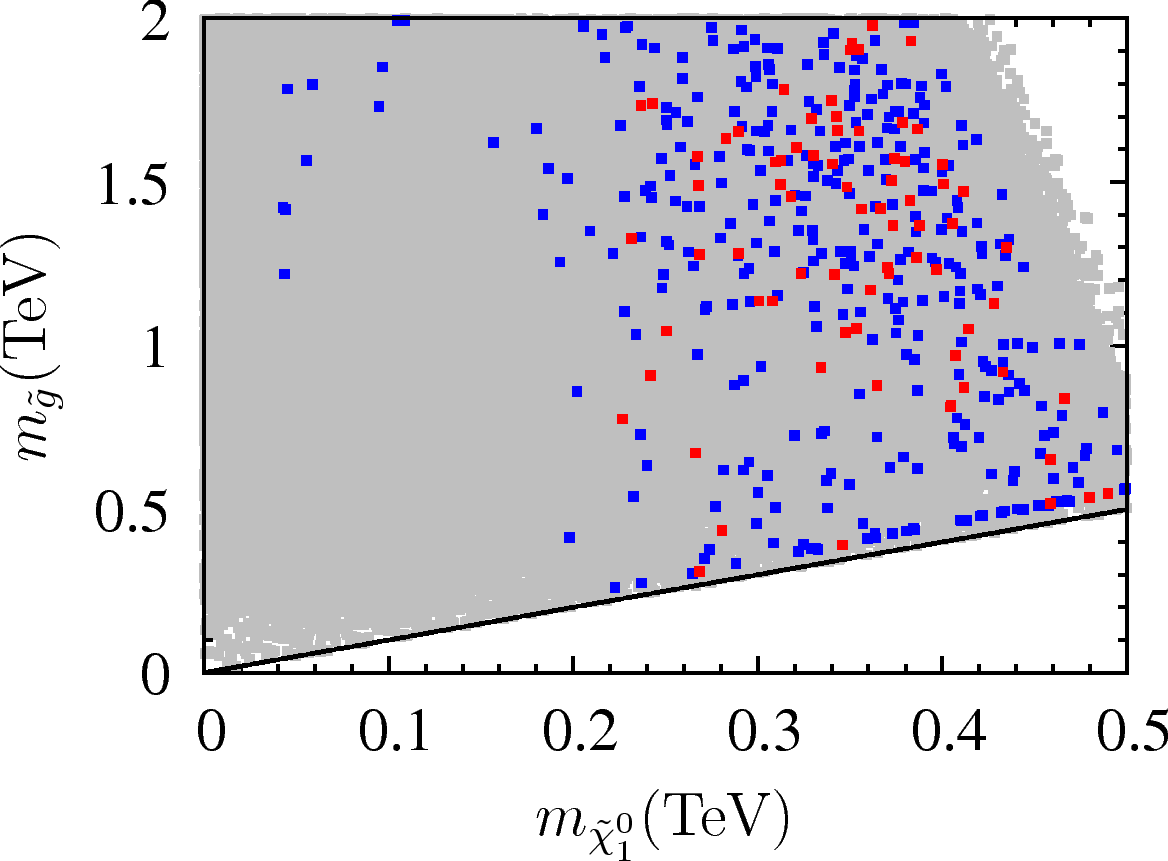}
}
\subfigure{
\includegraphics[width=7.2cm]{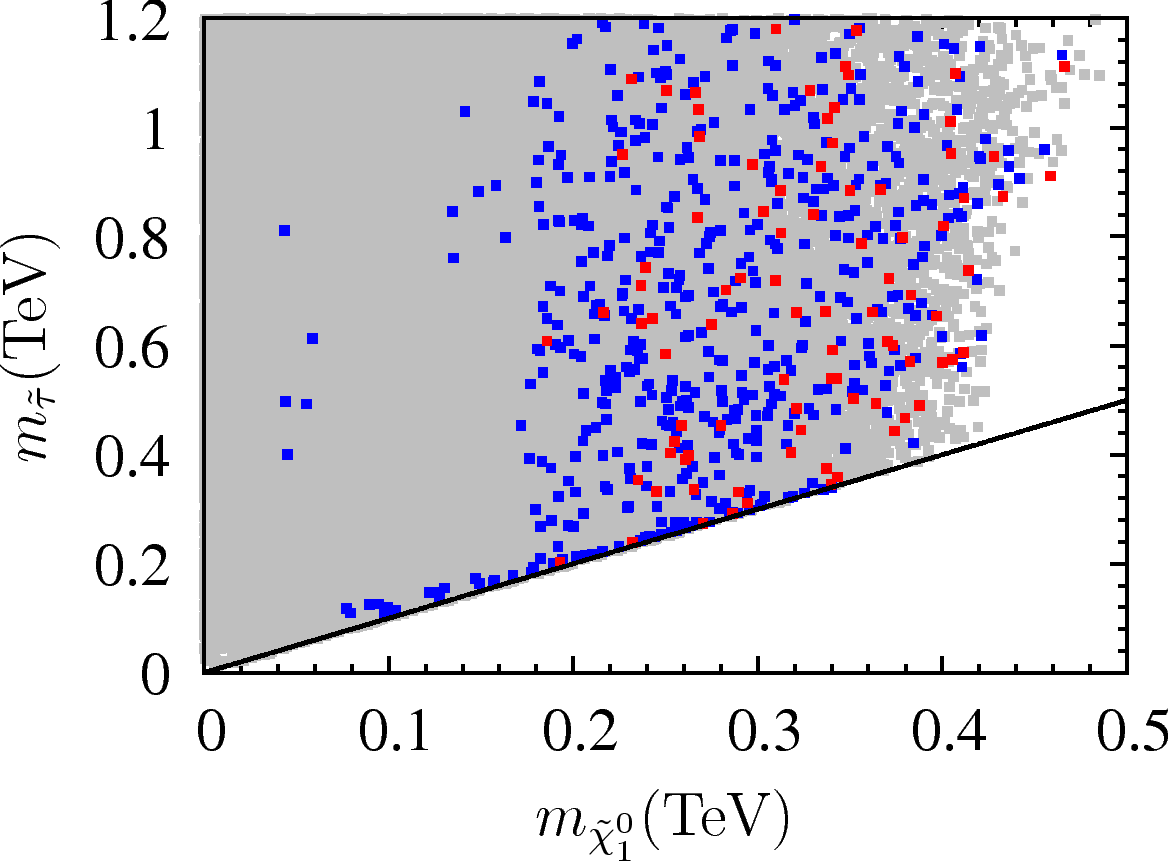}
}
\subfigure{
\includegraphics[width=7.2cm]{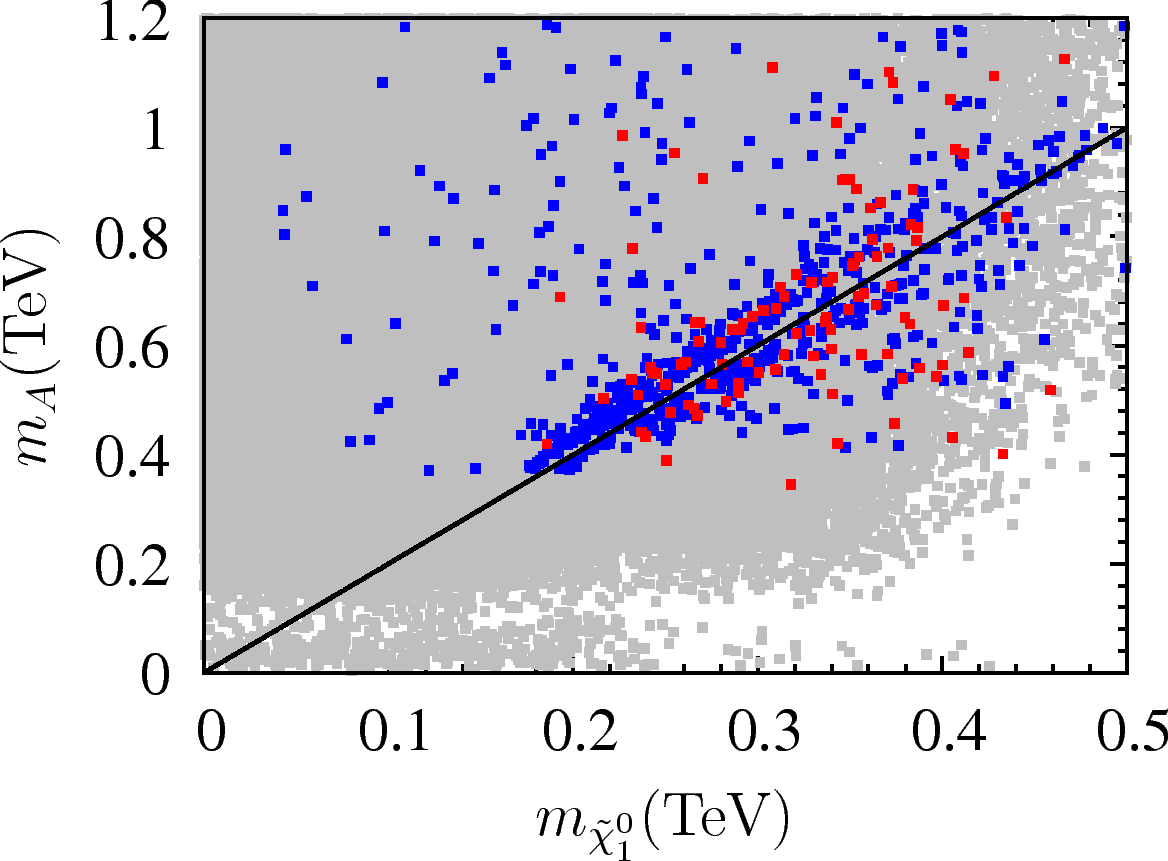}
}
\caption{
Plots in the $m_{\tilde{\chi}_1^{\pm}}$ - $m_{\tilde{
\chi}_1^0}$, $m_{\tilde {g}}$ - $m_{\tilde{ \chi}_1^0}$, $m_{\tilde
{\tau}}$ - $m_{\tilde{ \chi}_1^0}$ and $m_A$ - $m_{\tilde{
\chi}_1^0}$ planes. The gray points satisfy the requirements of REWSB and
$\tilde{\chi}^0_{1}$ LSP. The blue points, in addition, satisfy  particle mass
bounds and constraints from
$BR(B_s\rightarrow \mu^+ \mu^-)$,
$BR(B_u\rightarrow \tau \nu_{\tau})$ and $BR(b\rightarrow s \gamma)$.
In addition, we require that these points do no worse than the SM in
terms of the $(g-2)_\mu$ prediction.
The red points correspond to GYU to within 10\% in addition to these constraints.
We show in the $m_{\tilde{\chi}_1^{\pm}}$ - $m_{\tilde{
\chi}_1^0}$, $m_{\tilde {g}}$ - $m_{\tilde{ \chi}_1^0}$ and $m_{\tilde
{\tau}}$ - $m_{\tilde{ \chi}_1^0}$ planes the unit slope lines that
indicate the respective coannihilation channels. In the
$m_A$ - $m_{\tilde{\chi}_1^0}$ plane we show the line
$m_A=2m_{\tilde{\chi}_1^0}$ that signifies the $A$ resonance
channel.
\label{spectra}
}
\end{figure}

In Figure~\ref{GYm0tanbFig} we show the
results in the $GY$ - $m_0$ and $GY$ - $\tan\beta$ planes. There is
no visible distinction between the two classes of opposite sign
gaugino models in these two planes, which is why we plot the data from the
two sets in the same figure. The color coding is the same as in
Figure~\ref{M3M2Fig}. It can be seen that a relatively large
$m_0(\sim 500\,{\rm GeV})$ is required even without imposing any of the
experimental constraints. After including the experimental constraints, we are
forced to have $m_0\sim 1.5\,{\rm TeV}$. This is to be contrasted with
the situation depicted in Figure~\ref{compareModels} where 
$m_0\sim 300\,{\rm GeV }$ suffices for Yukawa coupling unification
compatible with all constraints. This may be understood from the fact
that keeping all other parameters fixed, a larger $m_0$ value tends to
push up the value of $y_t (M_{\rm GUT})$ closer to the
unified gauge coupling. Likewise, we must have
$46\lesssim \tan\beta \lesssim 54$ with an even narrower
range ($47\lesssim \tan\beta \lesssim 52$) if we consider 
the experimental constraints. Yukawa coupling unification, 
on the other hand, allows for $44\lesssim \tan\beta \lesssim 54$ 
for Yukawa unification consistent with experimental constraints.

In Figure~\ref{GYa0m0Fig} we show plots in the $GY$ - $A_0/m_0$
plane for the two classes of opposite sign gaugino models. The color
coding is the same as in Figure~\ref{M3M2Fig}. It is
evident that GYU with $\mu>0$ prefers $A_0/m_0<0$,
and vice versa. This is different from just Yukawa unified
4-2-2 as seen clearly from Figure~\ref{compareModels}. This stems from 
the finite chargino contribution to $\delta y_b$ which is
proportional to $\mu A_t$. In the case of Yukawa unification, one can have
a small $m_0$ value for which the chargino contribution is sub-dominant.
In GYU on the other hand, $m_0$ is
large as previously explained. This, coupled with the fact that we
need the threshold correction to $\delta y_b$ to be negative,
shows that $\mu A_0/m_0 <0$ is preferred for GYU.

In Figure~\ref{spectra} we show the relic density channels consistent
with GYU in the $m_{\tilde{\chi}_1^{\pm}}$ -
$m_{\tilde{ \chi}_1^0}$, $m_{\tilde {g}}$ - $m_{\tilde{ \chi}_1^0}$,
$m_{\tilde {\tau}}$ - $m_{\tilde{ \chi}_1^0}$ and $m_A$ -
$m_{\tilde{ \chi}_1^0}$ planes. The gray points in this figure
satisfy the requirements of REWSB and $\tilde{\chi}^0_{1}$ LSP. The blue points, 
in addition, satisfy the
particle mass bounds and constraints from $BR(B_s\rightarrow \mu^+
\mu^-)$, $BR(B_u\rightarrow \tau \nu_{\tau})$ and $BR(b\rightarrow s \gamma)$.
In addition, we require that these points do no worse than the SM in
terms of the $(g-2)_\mu$ prediction. 
The red points correspond to GYU to within 10\% in addition to these constraints.
We can see in Figure~\ref{spectra} that a variety of coannihilation
and annihilation scenarios are compatible with Yukawa unification
and neutralino dark matter. Included in the $m_A$ - $m_{\tilde{
\chi}_1^0}$ plane is the line $m_A$ = $2 m_{\tilde{ \chi}_1^0}$
which indicates that the $A$ funnel region is compatible
with GYU. In the remaining planes in
Figure~\ref{spectra}, we draw the unit slope line which indicates the
presence of gluino and stau coannihilation and bino-higgsino
mixed dark matter scenarios.

Recent results from ATLAS~\cite{Aad:2011hh,daCosta:2011qk} and 
CMS~\cite{Khachatryan:2011tk} naively put very stringent limits on 
the gluinop mass of $m_{\tilde g}\gtrsim 500\,{\rm GeV}$. However, it 
is shown explicitly in \cite{Akula:2011zq} that this limit 
does not apply in general, and specifically, does not 
apply in the case of heavy $\gtrsim \,{\rm TeV}$ squarks. In the 
gluino coannihlation channel shown in Figure~\ref{spectra} the 
squarks are heavy and are not yet excluded by ATLAS/CMS. The 
recent results do seem to suggest that the gluino coannihilation 
scenario will soon be tested.

\section{Gauge-Yukawa unification and dark matter detection\label{dark}}

\begin{figure}[t!]
\centering
\includegraphics[width=7.2cm]{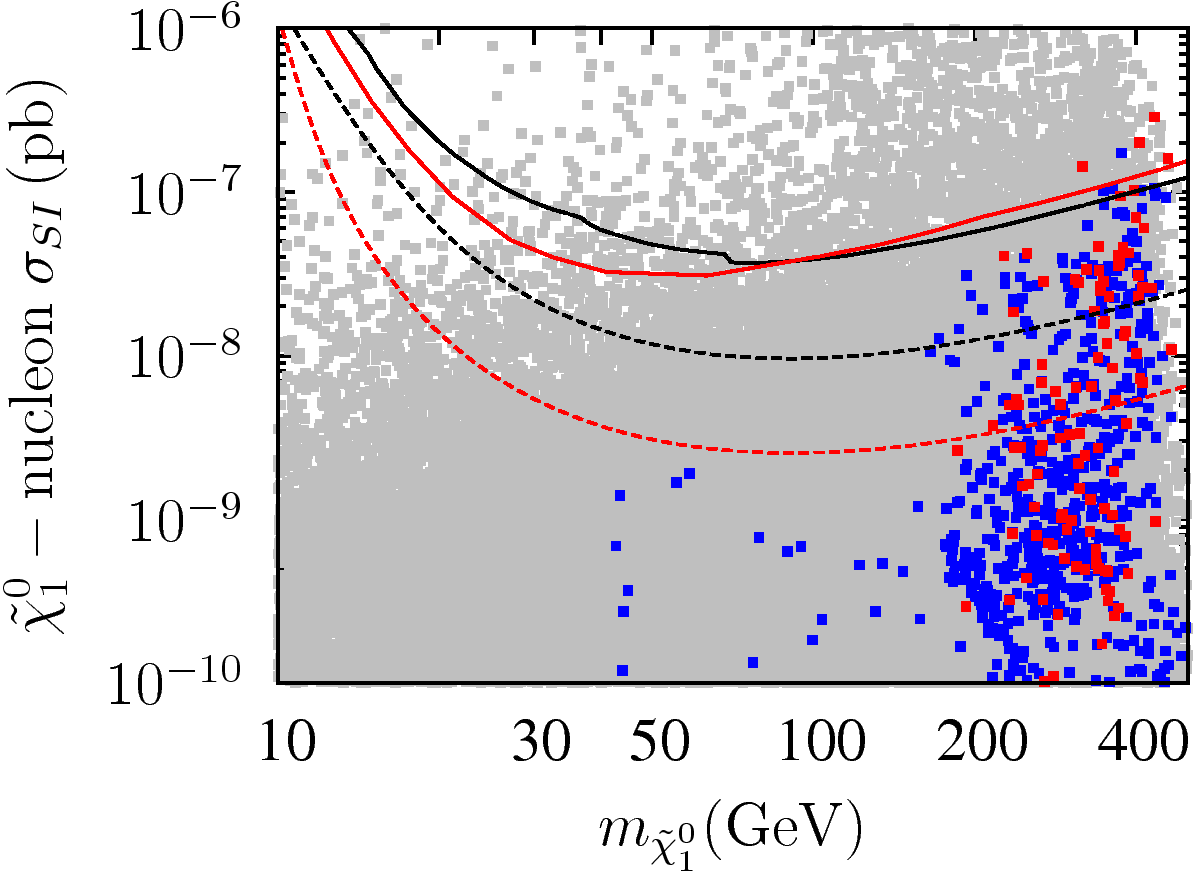}
\includegraphics[width=7.2cm]{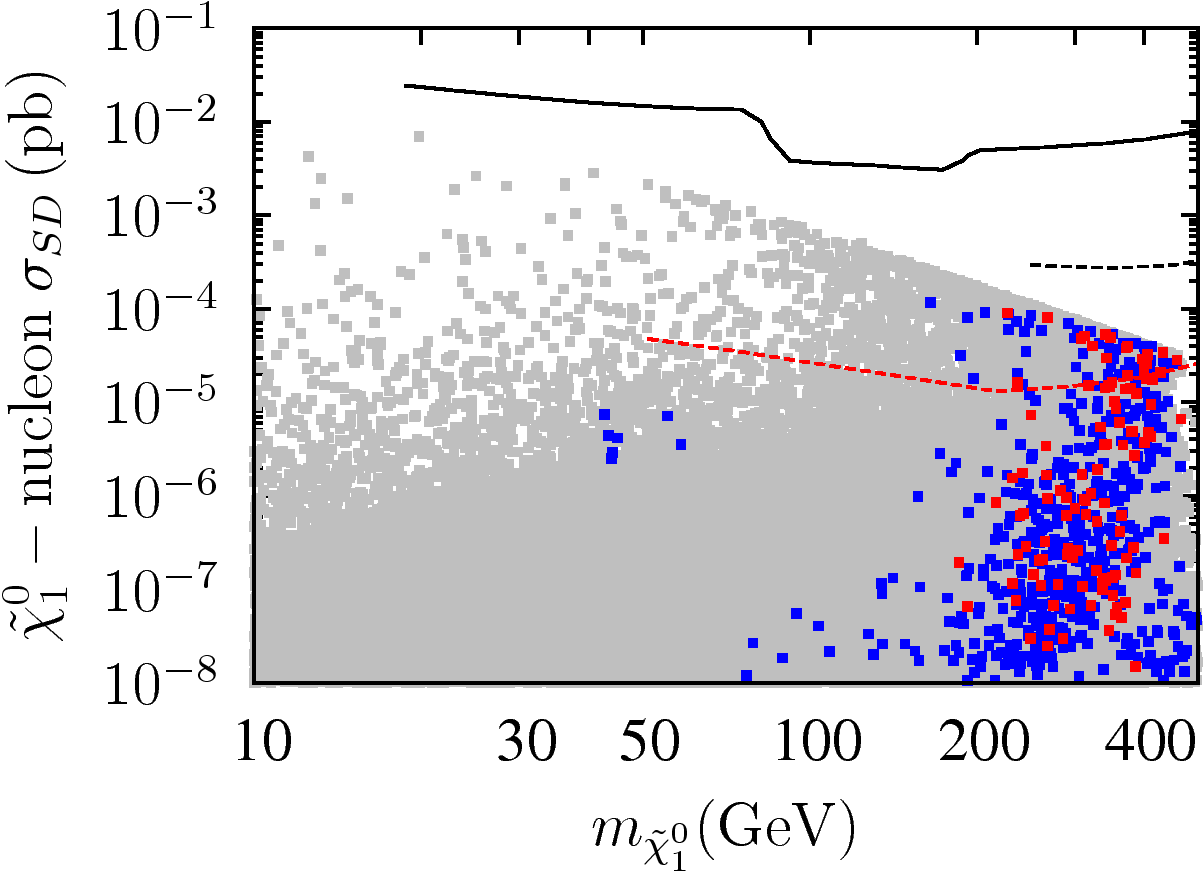}
\caption{Plots in the $\sigma_{\rm SI}$ - $m_{\tilde{\chi}_1^{0}}$
and $\sigma_{\rm SD}$- $m_{\tilde{\chi}_1^{0}}$ planes. Color coding
 is the same as in Figure~\ref{spectra}. In the $\sigma_{\rm SI}$ -
$m_{\tilde{\chi}_1^{0}}$ plane we show the current bounds (solid
lines) and future reaches (dotted lines) of the CDMS (black lines) and
Xenon  (red lines) experiments. In the $\sigma_{\rm SD}$ -
$m_{\tilde{\chi}_1^{0}}$ plane we show the current bounds from Super
K (black line) and IceCube (dotted black line) and future reach of
IceCuce DeepCore (dotted red line). \label{SIneuSDneu} }
\end{figure}

In light of the recent results by the CDMS-II \cite{Ahmed:2009zw}
and Xenon100 \cite{Aprile:2010um}  experiments, it is important to
see if GYU, within the framework presented in this
paper, is testable from the perspective of direct and indirect
detection experiments. The question of interest is whether $\mu \sim
M_1$ is consistent with GYU, as this is the
requirement to get a bino-higgsino admixture for the lightest
neutralino which, in turn, enhances both the spin dependent and spin
independent neutralino-nucleon scattering cross sections~\cite{Gogoladze:2010ch}.
In Figure~\ref{SIneuSDneu} we show the spin independent and spin
dependent cross sections as a function
of the neutralino mass. In the case of spin independent cross
section, we also show the current experimental bounds and future reach of the
CDMS and Xenon experiments. The color coding is the same as in
Figure~\ref{spectra}. A small region of the parameter space consistent
with GYU and the experimental constraints discussed
in Section~\ref{constraintsSection} (red points in the
figure) is at the exclusion limits set by the current CDMS and
XENON experiments. Thus, the ongoing and planned direct
detection experiments will play a vital role
in testing GYU models.

In the case of spin dependent cross section, we show in Figure~\ref{SIneuSDneu} the
current bounds from the Super-K \cite{Desai:2004pq} and IceCube
\cite{Abbasi:2009uz} experiments and the projected reach of IceCube
DeepCore. The current Super-K and IceCube bounds 
are not stringent enough to rule out anything. However,
from Figure~\ref{SIneuSDneu} we see that the future IceCube DeepCore
experiment will be able to constrain a significant region of the parameter space.

In Table~\ref{table1} we present some benchmark points for the
4-2-2 GYU model. All of these points are
consistent with neutralino dark matter and the constraints
mentioned in Section~\ref{constraintsSection}. Point 1 represents 
the best GYU that we have found and 
corresponds to the $A$ funnel region.
Points 2 and 3 correspond to the gluino and stau coannihilation 
channels, while for Point 4 bino-Higgsino mixing 
plays a major role in giving the correct dark matter relic density.  
As expected, both the spin independent 
and spin dependent cross sections of the neutralinos on protons 
are larger for Point 4. Note that Point 3 
also satisfies the lower bound on $\Delta (g-2)_\mu$. Finally, 
point 5 represents the best GYU solution that 
we found in the case of same sign gauginos ($GY\simeq1.14$).  
All of the points shown in this Table are currently allowed 
by ATLAS/CMS~\cite{Akula:2011zq}. 

\begin{table}[h!]
\centering
\begin{tabular}{lccccc}
\hline
\hline
                 & Point 1 & Point 2     & Point 3 & Point 4 & Point 5   \\
\hline
$m_{0}$          & 2063  & 3246    & 729   & 1769  & 7171  \\
$M_{1} $         & 747   & 1034    & -418  & 985   & 583   \\
$M_{2} $         & 1742  & 1819    & -1455 & 1938  & 939   \\
$M_{3} $         & -744  & -143    & 1138  & -443  & 49    \\
$\tan\beta$      & 50    & 48      & 47    & 50    & 53    \\
$A_0/m_0$        & -1.73 & 1.94    & -0.19 & -0.61 & -2.53 \\
$m_{Hu}$         & 2191  & 1162    & 657   &  1328 & 4557  \\
$m_{Hd}$         & 2797  & 3286    & 1294  &  2330 & 6722  \\
$m_t$            & 174.3 & 174.1   & 174.2 & 174.4 & 173.1 \\
sgn $\mu$        & +1    & +1      & -1    & +1    & +1    \\

\hline
$m_h$            & 117  & 119      & 119   & 116   & 121 \\
$m_H$            & 597  & 1739     & 694   & 1102  & 987 \\
$m_A$            & 594  & 1728     & 689   & 1095  & 983 \\
$m_{H^{\pm}}$    & 605  & 1742     & 701   & 1106  & 993 \\

\hline
$m_{\tilde{\chi}^0_{1,2}}$
                 &340, 528    &459, 1530  & 193, 1040 & 428, 492 &296, 907 \\
$m_{\tilde{\chi}^0_{3,4}}$
                 &529, 1492 &2708, 2710  &1058, 1256 &503, 1629 &6694, 6694 \\

$m_{\tilde{\chi}^{\pm}_{1,2}}$
                 &534, 1478  &1531, 2709  &1050, 1247 &497, 1618 &909, 6686\\
$m_{\tilde{g}}$  & 1750     & 516        & 2503 & 1128 & 340 \\

\hline $m_{ \tilde{u}_{L,R}}$
                 &2732, 2485  & 3445, 3209  &2451, 2260 &2319, 1946 &7186, 7110   \\
$m_{\tilde{t}_{1,2}}$
                 & 1355, 1793  & 1788, 2091  & 1846, 2091 &1153, 1643 &1948, 2607  \\
\hline $m_{ \tilde{d}_{L,R}}$
                 &2733, 2511 & 3445, 3276    & 2452, 2272 &2321, 1982 &7186, 7195 \\
$m_{\tilde{b}_{1,2}}$
                 &1336, 1781 &1485, 2065    &1801, 2074 &924, 1635   &2407, 2852 \\
\hline
$m_{\tilde{\nu}_{1}}$
                 & 2335       & 3412         & 1177   & 2148  & 7161  \\
$m_{\tilde{\nu}_{3}}$
                 &  1841      & 2814         & 1048  & 1835   & 5218 \\
\hline
$m_{ \tilde{e}_{L,R}}$
                &2336, 2115   & 3412, 3335  &1181, 784 & 2149, 1854  & 7160, 7251 \\
$m_{\tilde{\tau}_{1,2}}$
                & 540, 1836  & 1911, 2815 & 202, 1059 & 947, 1833 & 2129, 5204\\
\hline

$\sigma_{SI}({\rm pb})$
                & $9.1\times 10^{-9}$ & $4.7\times 10^{-12}$
                & $2.7\times 10^{-10}$ & $2.5\times 10^{-8}$
                & $1.1\times 10^{-12}$                          \\

$\sigma_{SD}({\rm pb})$
                & $5.6 \times 10^{-6}$ & $5.4 \times 10^{-10}$
                & $9.0\times 10^{-8}$ & $3.4\times 10^{-5}$
                & $7.8\times 10^{-12}$                          \\

$\Omega_{CDM}h^2$
                &  0.09      & 0.1     & 0.11 & 0.08  & 0.10  \\

$R$             &  1.05       & 1.07  & 1.08  & 1.04  & 1.13 \\
$GY$            &  1.05       & 1.09  & 1.09  & 1.06  & 1.14 \\
\hline
\hline
\end{tabular}
\caption{ 
Point 1 is the best GYU we found 
corresponding to the $A$ funnel region.
Points 2 and 3 respectively correspond to the gluino and stau coannihilation
channels, while for Point 4 the LSP is a bino-Higgsino admixture. Point 5 represents the
best GYU solution we found with same sign gauginos and
corresponds to gluino NLSP.
\label{table1}}
\end{table}

\section{Conclusions\label{conclusions}}

Guage-Yukawa unification (GYU) at $M_{\rm GUT}$, a natural extension
of four dimensional gauge unification, is implemented using
higher dimensional theories in which the gauge and
third ($t$-$b$-$\tau$) family matter supermultiplets are
unified. One of the simplest realizations of this idea gives
rise, after compactification, to the well-known
symmetry group $SU(4)_c \times SU(2)_L \times SU(2)_R$. GYU
in this framework strongly prefers gaugino
masses $M_2$ and $M_3$ with opposite signs, and it also shows
some preference for a top mass that is slightly higher than
$173.3$ GeV, its current central value. We have explored the
fundamental parameter space of GYU models and identify a
number of benchmark points that are compatible with a large
variety of experimental constraints, including the WMAP bound
on neutralino dark matter and $(g-2)_{\mu}$. One of the more
intriguing GYU compatible solutions corresponds to the
gluino NLSP scenario which can be tested at the LHC.

\section*{Acknowledgments}

This work
is supported in part by the DOE Grant No. DE-FG02-91ER40626
(I.G., S.R. and Q.S.), GNSF Grant No. 07\_462\_4-270 (I.G.) 
and the University of Delaware Competitive Fellowship (R.K.).

\end{document}